\documentclass[10pt]{article}
\usepackage{amsmath}
\usepackage{amsthm}
\usepackage{amsfonts}
\usepackage{amssymb}
\usepackage{url}
\usepackage{eucal}
\usepackage{mathrsfs}
\usepackage{float}
\usepackage{graphicx,graphics}
\usepackage{subcaption}
\usepackage[T1]{fontenc}%
\usepackage{url}
\usepackage{epsfig}
\usepackage{epstopdf}
\usepackage{xcolor}
\usepackage[english]{babel}

\newcommand\fig[1] {{\rm Figure}~\ref{fig:#1}}
\newcommand\figs[1] {~\ref{fig:#1}}
\newcommand\labfig[1] {\label{fig:#1}}

\newcommand{\bfm}[1]{\mbox{\boldmath ${#1}$}}
\newcommand{\nonum}{\nonumber \\}
\newcommand\eq[1] {(\ref{#1})}
\newcommand{\beqa}{\begin{eqnarray}}
\newcommand{\eeqa}[1]{\label{#1}\end{eqnarray}}
\newcommand{\beq}{\begin{equation}}
\newcommand{\eeq}[1]{\label{#1}\end{equation}}

\newcommand{\Grad}{\nabla}
\newcommand{\Div}{\nabla \cdot}

\newcommand{\Md}{\partial}

\newcommand{\Ga}{\alpha}
\newcommand{\Gb}{\beta}
\newcommand{\Gd}{\delta}

\newcommand{\Gg}{\gamma}
\newcommand{\Gc}{\chi}

\newcommand{\BGs}{\bfm\sigma}

\newcommand{\CP}{{\cal P}}

\newcommand{\CT}{{\cal T}}

\newcommand{\bpm}{\begin{pmatrix}}
\newcommand{\epm}{\end{pmatrix}}

\def\b0{\bf 0}

\def\Be{{\bf e}}

\def\Bj{{\bf j}}

\def\Bn{{\bf n}}

\def\Bx{{\bf x}}

\usepackage{caption,subcaption}
\begin{document}
\vspace{-1in}
\title{Field patterns without blow up}
\author{Ornella Mattei and Graeme W. Milton\\
	\small{Department of Mathematics, University of Utah, Salt Lake City UT 84112, USA}}
\date{}
\maketitle
\vskip 1.cm
\begin{abstract}
	\noindent
Field patterns, first proposed by the authors in [G. W. Milton and O. Mattei, Proc. R. Soc. A 473, 20160819 (2017)], are a new type of wave propagating along orderly patterns of characteristic lines which arise in specific space-time microstructures whose geometry in one 
spatial dimension plus time is somehow commensurate with the slope of the characteristic lines. In particular, in [G. W. Milton and O. Mattei, Proc. R. Soc. A 473, 20160819 (2017)] the authors propose two examples of space-time geometries 
in which field patterns occur: they are two-phase microstructures in which rectangular space-time inclusions of one material are embedded in another material. After a sufficiently long interval of time, field patterns have local periodicity both in time and space. This allows one to focus only on solving 
the problem on the discrete network on which a field pattern lives and to define a suitable transfer matrix that, given the solution at a certain time, provides the solution after one time period. For the aforementioned microstructures, many of the eigenvalues of this $\CP\CT$--symmetric 
transfer matrix have unit norm and hence the corresponding eigenvectors correspond to propagating modes. However, there are also modes that blow up exponentially with time coupled with modes 
that decrease exponentially with time. The question arises as to whether there are space-time microstructures such that the transfer matrix only has eigenvalues on the unit circle, so that there are 
no growing modes (modes that blow-up)? This finds answer here, where we see that certain space-time checkerboards have the property that 
all the modes are propagating modes, within a certain range of the material parameters. Interestingly, when there is no blow-up, the waves generated by an instantaneous 
disturbance at a point look like shocks with a wake of oscillatory waves, whose amplitude, very remarkably, does not tend to zero away from the wave front.

\end{abstract}

\section{Introduction}
\setcounter{equation}{0}

Field patterns \cite{Milton:2017:FP} are a new sort of wave that occur in space-time microstructures with a geometry chosen such that a disturbance
does not generate a complicate cascade of subsequent disturbances, but rather concentrates on a periodic space-time pattern,
that we call the field pattern. The examples of field patterns discovered by the authors in \cite{Milton:2017:FP} support both
propagating waves and also waves that can blow-up exponentially with time, or decay exponentially with time. This paper
presents the first examples of space-time microstructures that support field patterns such that all the modes are propagating modes, 
with no blow-up.

The study of space-time materials and, more specifically, of space-time microstructures was initiated in the late nineteen 
fifties \cite{Cullen:1958:TWP,Louisell:1958:PAS,Morgenthaler:1958:VME,Tien:1958:PAF,Honey:1960:WBU}
and has seen a recent surge of interest \cite{Lurie:2006:WPE,Lurie:2009:MAW,To:2009:HDL,Yu:2009:COI,Rousseau:2011:ESD,Sanguinet:2011:HEM,Lira:2012:EDN,Fang:2012:REM,Taravati:2016:MDA,Yuan:2016:PGP}:
see, for example, the book \cite{Lurie:2007:IMT} and the introduction in \cite{Milton:2017:FP} for a more comprehensive review of the relevant literature, as well as 
references to ways in which space-time microstructures can be generated.
Space-time microstructures represent composites whose material properties vary not just with respect to space, like standard composites, but also with respect to time. 
In particular, space-time microstructures can be divided into two groups depending on how the properties of the material are changed in time: the so-called activated 
materials are realized by an external mechanism that produces a time switching of the space pattern of the material in a pre-determined manner, whereas the so-called 
kinetic materials are realized by an actual mechanical motion of the various parts of the composite system with respect to the laboratory frame (see \cite{Lurie:2007:IMT}). 

Wave propagation in space-time geometries can have some interesting properties beyond the occurence of field patterns.  
For instance, by suitably controlling the design parameters of a space-time activated laminate it is possible to selectively screen large space-time domains from long wave disturbances, see \cite{Lurie:1997:EPS,Weekes:2001:NCW}; or, if the parameters of the constituents of a space-time checkerboard are suitably chosen, one has, within each space period, disturbances converging towards a so-called ``limit cycle'' after a few time periods and when this happens the energy in the system grows exponentially (see, e.g., \cite{Lurie:2016:EAW}). In general, every time 
a wave hits a space-time boundary it splits into two waves and consequently a single disturbance usually evolves into a complicated cascade of disturbances. Thus keeping track of how 
the wave propagates through the space-time microstructure leads to complex numerical simulations. With the aim of overcoming such a disadvantage, and in the process of revealing 
dramatically new wave
behavior, the authors introduced in \cite{Milton:2017:FP} the theory of field patterns: the space-time microstructure is chosen in such a way that its geometry is, in a sense, related to the slope of the characteristic lines so that the disturbance does not evolve into a cascade of disturbances but rather concentrates on a pattern of characteristic lines: this is the field pattern. In particular, the space-time geometries giving rise to field patterns that were proposed by the authors in \cite{Milton:2017:FP} are two-phase composites made of space-time rectangular inclusions of a certain material embedded in a space-time matrix made of another material. The placement of the rectangular inclusions is such that the space-time geometry is endowed with both $\CP$ symmetry and $\CT$--symmetry, where $\CP$--symmetry refers to reflection invariance of the microstructure under the parity operation of spatial reflection $x\rightarrow - x$ when the origin is chosen at the center of an inclusion, and $\CT$-symmetry stands for time symmetry, the reflection invariance of the microstructure under time reversal $t\rightarrow -t$ (when again the origin is chosen at the center of an inclusion). 

The key idea of studying the dynamics of systems with $\CP\CT$--symmetry was  made by Bender and coauthors in the context of quantum physics (see, e.g., the pioneering paper \cite{Bender:1998:SSD} by Bender and Boettcher). They proved that 
the Hermiticity condition of the Hamiltonian, guaranteeing that the time evolution is unitary, is only sufficient but not necessary to have a real spectrum. Under the weaker condition of $\CP\CT$--symmetry, where the system is
invariant under the combined operation of spatial reflection and time reversal,  the spectrum can be real and positive below a certain threshold, above which the eigenvalues become complex conjugates of each other (see also \cite{Bender:2007:MSN} for a comprehensive review).
Due to the isomorphism between the Schr{\"o}dinger equation and the paraxial Helmholtz equation, $\CP\CT$--symmetry has been extended to optical and plasmonic systems, with the role of the potential in quantum mechanics replaced by the refractive index of the material (see, e.g., the review article \cite{Zyablovsky:2014:PTS}). In particular, the $\CP\CT$--symmetry condition implies that the real part of the permittivity coefficient is an even function of position while its imaginary part is odd. When the optical eigenvalues are complex, certain eigenmodes will experience increased loss whereas other eigenmodes will exhibit strong optical gain, thus leading to the asymmetric and unidirectional optical properties observed in $\CP\CT$--metamaterials (see, for instance, \cite{Klaiman:2008:VBP,Makris:2008:BDI,Kottos:2010:BSM,Ruter:2010:OPT,Alaeian:2014:PTS,Savoia:2015:PTS}).

As a consequence of the special geometry of the space-time microstructures presented in \cite{Milton:2017:FP}, endowed with both 
$\CP$ and $\CT$--symmetry, the network of characteristic lines 
on which the field pattern lives is periodic both in space and time. An individual field pattern has 
$\CP\CT$--symmetry, but breaks the separate symmetries 
under spatial reflection and time reversal of the underlying space-time microgeometry. The network nature
of field patterns allows one to discretize the problem. Everything boils down to the calculation of a discrete Green function,
involving one cell and its neighbours, which enables one to generate an associated transfer matrix that allows one to step through time. The transfer matrix relates the solution on the discrete network at a certain time to the solution on the discrete network after one time period. In general, the transfer matrix has determinant equal to one (due to the $\CP\CT$--symmetry) and its eigenvalues could have either modulus one (in this case the corresponding mode is a propagating mode), or modulus greater than one, thus corresponding to modes that blow up exponentially with time (due to the $\CP\CT$--symmetry, these eigenvalues are coupled with eigenvalues with modulus less than one, corresponding to modes that decrease exponentially with time): it is the choice of the initial conditions that, within the same field pattern, leads to either a propagating or an increasing (or decreasing) mode.

Besides opening new avenues of research regarding wave propagation in space-time microstructures, the theory of field patterns is particularly interesting also for some of its features which hint at a possible connection with quantum mechanics: field patterns exhibit both a particle-like aspect and a wave-like aspect, as they are propagating waves concentrated on lines in a space-time diagram.
Furthermore, even though they appear in scalar wave equations, they have a multidimensional or multicomponent nature. In fact, in the linear model developed in \cite{Milton:2017:FP}, field patterns with different spatial shifts do not interact, but rather evolve as if they lived in separate dimensions (as many dimensions as the number of field patterns), 
with each dimension playing the same role of the others, as do the different dimensions associated with say the multielectron Schr{\"o}dinger equation with noninteracting electrons.

Moreover, the theory of field patterns could be very useful in the study of spatially periodic 
composites of hyperbolic materials. We recall that a material is called hyperbolic (see, e.g., \cite{Fisher:1969:RCF}) if the dielectric tensor has both positive and negative eigenvalues, with at least the negative one having a small imaginary part that represents loss and 
causes absorption of electromagnetic energy into heat. The important feature about these materials is that they can direct radiation along the ``characteristic lines'' in the hyperbolic medium, thus giving rise, for specific configurations, to the so-called hyperlenses (see, for instance, \cite{Jacob:2006:OHF,Salandrino:2006:FFS,Rho:2010:SHT,Lu:2012:HMF}). In the limit where the loss goes to zero, a two-dimensional spatially periodic hyperbolic material can be viewed as a space-time microstructure in one spatial dimension plus time, if one reinterprets one of the two spatial coordinates as time. Therefore, the results obtained within the theory of field patterns may be applicable to hyperbolic materials in the limit where the imaginary part of 
the dielectric tensor goes to zero. Of course the physics should dictate which modes are
relevant in this limit (in an infinite two-phase hyperbolic material, with no defects, 
one does not expect the exponentially growing modes or exponentially decaying modes to have a physical significance).
It remains to determine the conditions under which a solution in a hyperbolic material with non-zero loss converges to a solution of the equations with zero loss, as the loss 
(imaginary part of the dielectric tensor) goes to zero.

Among all the research avenues launched by the theory of field patterns that still have to be explored, here we focus on finding space-time microstructures, in which field patterns occur, 
such that there are no modes which blow-up in time. In other words, we seek field patterns such that the transfer matrix has all its eigenvalues on the unit circle, so that there are
only propagating modes. In particular, we achieve such a goal by considering space-time multi-component checkerboards in which the wave speeds of the constituents satisfy special relations.

We find some curious features associated with the waves generated by a single, instantaneous, point disturbance, i.e., with the Green function, that can be seen in figures
\figs{check1_inj1}, \figs{check4_prop} and \figs{check3phases_inj1}. It is almost as if a shock is generated leaving a wake behind it of oscillatory waves and very interestingly, the amplitude of the oscillations does not tend to zero as one goes away from the wave-front. This seems vaguely
similar to the way a real shock heats air and leaves in its wake thermal oscillations. However, in contrast to real shocks where energy in the shock is transferred to heat, the magnitude of the "shock" in a field pattern wave does not decrease in time".  

\section{The fundamentals of the theory of field patterns}
\setcounter{equation}{0}

The theory of field patterns can be formulated in various ways that reduce to a wave equation in a space-time geometry. Here we find it mathematically convenient to use a formalism close to that of the electric conductivity problem, that is also the point of view considered in 
\cite{Milton:2017:FP}. Let us focus then on the two-dimensional conductivity equation
\beq \Bj(\Bx)=\BGs(\Bx)\Be(\Bx),\quad{\rm where}\quad \Div\Bj=0, \quad \Be=-\Grad V, \eeq{1.1}
with $\Bj(\Bx)$ the electric current, $\Be(\Bx)$ the electric field, $V(\Bx)$ the electric potential, and $\BGs(\Bx)$ the conductivity tensor. We consider a $p$-component composite, where the phases have conductivities $\BGs_i$, $i=1,\dots,p$ so that $\BGs(\Bx)$ can be written as
\beq \BGs(\Bx)=\sum_{i=1,p}\Gc_i(\Bx)\BGs_i, \eeq{1.2}
where $\Gc_i(\Bx)$ is the characteristic function of phase $i$ (equal to 1 if $\Bx$ belongs to phase $i$ and zero otherwise), and we suppose that all the $\BGs_i$, with $i=1,\dots,p$, take the form
\beq \BGs_i=\bpm \Ga_i & 0 \\ 0 & -\Gb_i \epm\eeq{1.3}
where the parameters $\Ga_i$ and $\Gb_i$, $i=1,\dots,p$ are, in general, real and positive. 
Note that if $\BGs(\Bx)$ and  $\BGs_i$, with $i=1,\dots,p$, were dielectric tensors and the coefficients $\Ga_i$ and $\Gb_i$, $i=1,\dots,p$, were complex-valued functions with non-negative imaginary part, then the material under study would be a hyperbolic material.  

The global potential $V(\Bx)$ can be written as 
\beq V(\Bx)=\sum_{i=1,p}\Gc_i(\Bx)V_i(\Bx), \eeq{V_global}
$V_i(\Bx)$ being the potential in phase $i$, $i=1,\dots,p$, fulfilling within that phase the following wave equation, obtained by combining equations \eq{1.1} and \eq{1.3}:
\beq \Ga_i\frac{\Md^2 V_i}{\Md x_1^2}= \Gb_i\frac{\Md^2 V_i}{\Md x_2^2} \eeq{1.4}
where we can think of $x_1$ as representing the space variable $x$, and $x_2$ as representing the time variable $t$. This corresponds to considering a one-dimensional space plus time distribution of $p$ materials, each having conductivity coefficient in the space direction equal to $\Ga_i$ and conductivity coefficient in the time direction equal to $-\Gb_i$, $i=1,\dots,p$. 
The D'Alembert solution of equation \eq{1.4} gives the local (not global) solution in phase $i$ as the sum of two independent waves:
\beq V_i(x,t)=V^+_i(x-c_it)+V^-_i(x+c_it) \eeq{1.5}
with $V^+_i(x-c_it)$ the wave moving upwards to the right in a space-time diagram and $V^-_i(x+c_it)$ the wave moving upwards to the left, with the wave speed $c_i$ defined as
\beq c_i=\sqrt{\Ga_i/\Gb_i}. \eeq{1.6}

To determine the explicit expression of the potentials $V_i(x,t)$, $i=1,\dots,p$, and therefore the solution of the problem in terms of the global potential $V(x,t)$, we need to choose the initial conditions (for what concerns the boundary conditions, we suppose the medium to be infinite in the $x$ direction). Let us suppose that $V(x,t)$ satisfies the following initial conditions: 
\beq V(x,0)=v_0H(x-a), \quad j_t(x,0)=\Gd(x-a)j_0, \eeq{1.7a}
with $H(y)$ the Heaviside function
and $\Gd(y)$ the Dirac delta function. Thus, at time $t=0$, we are injecting, in the time direction ($j_t$ is the $t$-component of $\Bj$), a total current flux $j_0$ concentrated at $x=a$, in correspondence to which there is a jump in potential. Such a current flux will split along the two characteristic lines starting at the point $(x,t)=(a,0)$ and having slope respectively equal to  $c_k$ and $-c_k$, with $k$ the phase the point $(a,0)$ belongs to. If the initial jump in potential, $v_0$, is equal to $v_0={j_0}/{c_k\,\Gb_k}$, then the potentials $V^+_k(x,t)$ and $V^-_k(x,t)$ in phase $k$ will take the following expressions:
\beq V^+_k(x,t)=s_k^+[1-H(x-c_kt)],\quad V^-_k(x,t)=s_k^-H(x+c_kt), \eeq{1.9} 
with $s_k^+=-{j_0}/(c_k\Gb_k)=-s_k^-$, and the currents flowing along the characteristic lines in phase $k$ will be
\beqa  \Bj^+_k=s_k^+\sqrt{\Ga_k\Gb_k}\bpm c_k \\ 1 \epm \Gd(x-c_kt)\equiv \frac{s_k^+}{\Gg_k}\frac{1}{\sqrt{1+c_k^2}}\bpm c_k \\ 1 \epm \Gd(x-c_kt), \nonum
\Bj^-_k=s_k^-\sqrt{\Ga_k\Gb_k}\bpm -c_i \\ 1 \epm \Gd(x+c_kt)\equiv \frac{s_k^-}{\Gg_k}\frac{1}{\sqrt{1+c_k^2}}\bpm -c_k \\ 1 \epm \Gd(x-c_kt),
\eeqa{1.10}
with 
\beq \Gg_k=\frac{1}{\sqrt{\Ga_k(\Ga_k+\Gb_k)}}. \eeq{1.11}
Therefore, in phase $k$ the potential is piecewise constant and the amplitude of the potential jump across a certain characteristic line is given by the product of the coefficient $\Gg_k$ and the modulus of the current flowing through that characteristic.

Now, when the disturbance, moving along the characteristic lines in phase $k$, hits a space-time interface, it splits into two waves and 
as demanded by the weak form of \eq{1.1}
we suppose that, when this happens, the continuity of the potential and the continuity of the current flux are preserved, that is
\beqa V_k(x,t)=V_l(x,t)
\label{continuity_V}\\ \Bn\cdot\BGs_k\Grad V_k(x,t)=\Bn\cdot\BGs_l\Grad V_l(x,t)\,,\eeqa{Continuity_flux}
with $\Bn^{\mbox{\footnotesize{T}}}=(n_x\,\,n_t)$ being the normal vector to the interface between phase $k$ and phase $l$, and $\Grad^{\mbox{\footnotesize{T}}}=(\partial/\partial x\,\,\,\,\partial/\partial t)$. 
In general, if the space-time boundaries between phases are not suitably placed, the disturbances will branch giving rise to a complicated cascade of current lines 
that is difficult to analyze. However, when the geometry of the space-time microstructure is chosen in an appropriate way (see \fig{check1} as an example), the characteristic lines will form an orderly pattern, called a field pattern \cite{Milton:2017:FP}, with a locally periodic behavior in space and time. 
For instance, in \cite{Milton:2017:FP}, the authors proposed two types of space-time geometries composed of rectangular inclusions of material 2 immersed in material 1, with the size of the rectangles and the distance between them chosen in relation to the slopes of the characteristics. Both geometries, one presenting aligned inclusions and the other one with staggered inclusions, are examples of microstructures endowed with $\CP\CT$--symmetry. We point out that these space--time microstructures (as all the other geometries presented in the following, see Sections \ref{Check_same_speed} and \ref{Check_spec_ratio}) are all endowed with  $\CP$--symmetry and $\CT$--symmetry separately, if one chooses the origin of the axes in a space time--diagram to coincide with the center of one of the inclusions, for the geometries proposed in \cite{Milton:2017:FP}, or with the center of one of the squares composing the checkerboard, for the materials presented in Sections \ref{Check_same_speed} and \ref{Check_spec_ratio}. In the chosen coordinate system, the periodic pattern of characteristics associated with a single field pattern is $\CP\CT$--symmetric (but generally not separately $\CP$-- and $\CT$--symmetric). Note that $\CP\CT$--symmetric field patterns can occur in space--time microstructures without both $\CP$-- and $\CT$--symmetry (maybe only with $\CP\CT$--symmetry) as will be clear later from th examples in \fig{alternative_check}.

One of the key features of the theory of field patterns is that, since a field pattern lives on its own discrete network\footnote{We recall that, in general, the choice of a different $x$--coordinate of the launching point $(a,0)$ generates a different network of characteristic lines, and thus a different field pattern. If current is initially injected at multiple launching points, the field patterns that arise in the space--time microstructure will not in general interact but each will evolve on its own network as if living in a separate dimension, see \cite{Milton:2017:FP}.}, it is only
necessary to study the dynamics of field patterns on these discrete networks, that is, 
it suffices to study the current distribution along the characteristic lines at discrete moments of time, say $t=\tau+n\,t_0$ for $n=0,1,2,...$, where $t_0$ is the periodicity in the time direction of the discrete network and $\tau$ is some fixed time (for simplicity $\tau$ is chosen in such a way that none of the characteristic lines
intersect at $t=\tau$, see \fig{check1} as an example). Then, the state of the system is captured by the function $j(k,m,n)$, where the integer $k$ indexes the current line within the unit cell ($k$ is the number of intersection points between the characteristic lines and the horizontal line $t=\tau$ within a unit cell), the integer $m$ indexes the cell, and the integer $n$ indexes the discrete time.
To determine the evolution of the state function $j(k,m,n)$ as $n=0,1,2,...$ increases, one has to calculate the Green function that allows one to recover the currents at a certain time $t=\tau+n\,t_0$ with $n$ fixed, given the currents at time $t=\tau+(n-1)\,t_0$. We recall that the components of the Green function are determined by 
taking one unit cell, say with $m=m_0$ and injecting, at $t=\tau$ ($n=0$), a unit current in each of the $k$ intersecting points, one at a time, and by calculating how such a current flows along the characteristic lines to determine the currents at $t=\tau+t_0$ ($n=1$). We denote the Green function by $G_{k,k'}(m-m')$ to indicate that it provides the current at point $k$ of cell $m$, given the current at point $k'$ of cell $m'$. Such a function obviously only depends on the differences $m-m'$. Then, the current at the point $k$ of cell $m$ at time $t=\tau+n\,t_0$ is determined by the currents at points $k'$ of cells $m'$ at the previous time $t=\tau+(n-1)\,t_0$ by 
\beq j(k,m,n)=\sum_{k',m'}T_{(k,m),(k,'m')}j(k',m',n-1), \eeq{3.1}
where
\beq T_{(k,m),(k,'m')}=G_{k,k'}(m-m') \eeq{3.2}
is the $\CP\CT$--symmetric transfer matrix. 

The space-time microstructures presented in \cite{Milton:2017:FP}, although very interesting as they represent the first examples of geometries in which field patterns occur, are such that the associated transfer matrix has eigenvalues with modulus equal to one (giving rise to propagating modes) but also eigenvalues with modulus bigger than one (coupled with eigenvalues with modulus less than one, due to the $\CP\CT$-symmetry of the geometry), thus corresponding to modes that blow up (decrease) exponentially with time. In the following sections we will present examples of $\CP\CT$--symmetric space-time microstructures, in particular checkerboards, for which all the eigenvalues of the transfer matrix have modulus equal to one so that there are only propagating modes. We will see that for some of the microstructures analyzed in the following section, the condition of $\CP\CT$--symmetry is ``unbroken'' (all the eigenvalues lie on the unit circle) independently of the choice of the material parameters, whereas for other materials there will be sets of parameters leading to ``unbroken $\CP\CT$--symmetry'' and others yielding ``broken $\CP\CT$--symmetry'' (some of the eigenvalues lie on the unit circle and those that do not are complex conjugates of each other).

We point out that, in the numerical analyses presented in the next sections, to approximate the hypothesis of an infinite medium in the $x$ direction, we will consider periodic boundary conditions, so that we can think of the discrete network as lying on the surface of a cylinder, with the axis variable
corresponding to time and the angle variable corresponding to space. Thus, in \eq{3.1} cell $M$ is identified with cell $0$ and cell $-1$ is identified with
cell $M-1$: equivalently, the argument $m-m'$ of $G_{k,k'}(m-m')$ in \eq{3.1} should be replaced with $(m-m')\mod M$. 

We also remark that, to test numerically the $\CP\CT$--symmetry of the transfer matrix, one has first to reflect all the columns with respect to, say, the central column (due to the periodic boundary conditions any column can be considered as the reference column), so that, for instance, the entries of the first column become the entries of the last one and vice versa (this corresponds to applying the $\CP$--operator of space reflection). Then, one has to calculate the inverse of the matrix so obtained (this corresponds to applying the $\CT$--operator of time reversal). Finally, after reflecting again the columns of this new matrix with respect to its central column, one has to calculate the inverse (this operation corresponds to applying the $(\CP\CT)^{-1}$ operator). The result is a matrix that is equal to the original transfer matrix.

\section{Multi-component space-time checkerboards with phases having the same wave speed}\label{Check_same_speed}

To start with, we look for space--time microstructures that give rise to field patterns supporting only propagating modes by looking at the simplest non-trivial geometry, that is, the checkerboard geometry, already studied by Lurie and coworkers (see, for instance, \cite{Lurie:2009:MAW}). For simplicity, and to generate a microstructure that supports a field pattern, we consider a two-phase space-time checkerboard with the two components having the same wave speed, $c_1=c_2=c$ and with the diagonals of the rectangles composing the checkerboard having the same slope of the characteristic lines (see \fig{check1}, left). If we suppose the height of each rectangle to be unity, its width will be equal to $c$. The unit cell of the field pattern, periodic both in space and time, corresponds to the unit cell of the checkerboard geometry (see \fig{check1}, on the right). Obviously, the ratio between its space dimension $x_0$ and its time dimension $t_0$ is equal to $c$.  
 
\begin{figure}[!ht]
\includegraphics[width=\textwidth]{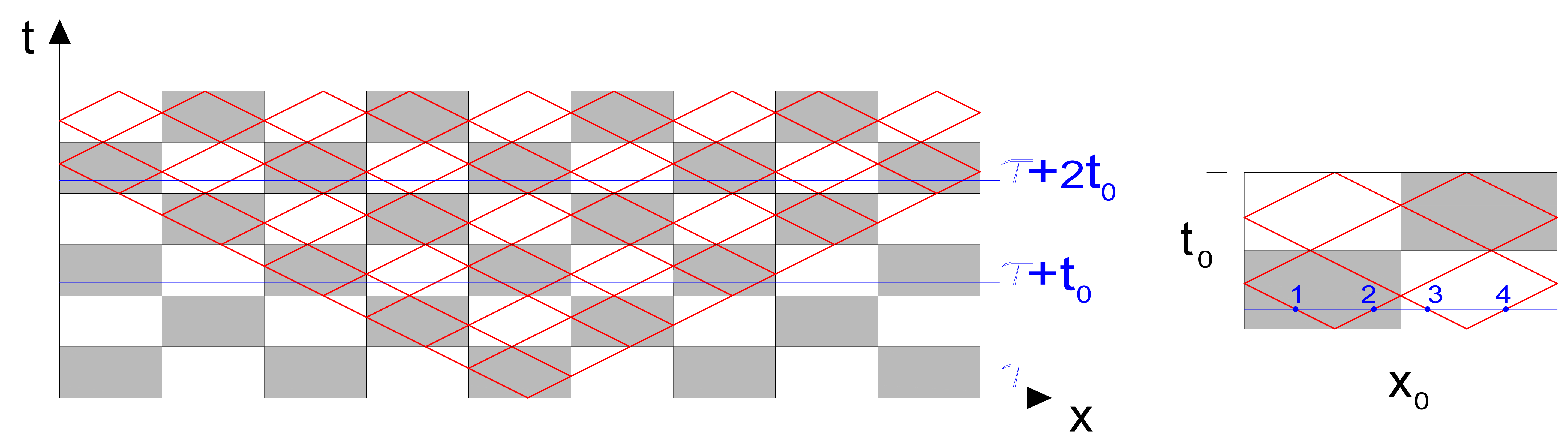}
\caption{On the left is a two-phase space-time checkerboard with the two components having the same wave speed, $c_1=c_2=c$, giving rise to a field pattern (the red lines represent the characteristic lines). In particular, phase 1 is the one colored in white and phase 2 the one colored in gray. On the right, the unit cell of the dynamic network, corresponding to the unit cell of the space-time geometry. The dimensions of the rectangles in the checkerboard are chosen such that the diagonals have the same slope of the characteristic lines, so that, if the time dimension of the unit cell is $t_0=2$, then its space dimension is $x_0=2c$. Due to the discrete nature of the field pattern and by virtue of its periodicity in time and space, we will evaluate the solution of the conductivity problem only at specific points on the discrete network at specific times. In particular, by choosing a certain time $t=\tau$, with the horizontal line $t=\tau$ intersecting, for instance, the characteristic lines in the four points within the unit cell indicated in the figure on the right, it is sufficient to evaluate the solution at the same four points of each unit cell at the subsequent times $t=\tau+nt_0$, with $n=1,2,\dots$.}
\labfig{check1}     	
\end{figure}

To determine the transfer matrix for this space-time geometry we follow the procedure introduced in \cite{Milton:2017:FP} and recalled at the end of the previous section. First, we fix a specific moment of time $\tau$ such that none of the current lines in the discrete network
intersect at $t=\tau$, as in \fig{check1}. In this case, the number of current lines intersected by the horizontal line $t=\tau$ in a unit cell is equal to 4 (the 4 points marked by blue dots in \fig{check1} on the right). Now, the components of the Green function are determined by 
taking one unit cell and injecting, at $t=\tau$ ($n=0$), a unit current in each of the $4$ intersecting points, one at a time, and by calculating how such a current flows along the characteristic lines to determine the currents at $t=\tau+t_0$ ($n=1$). Since the Green function $G_{k,k'}(m-m')$ only depends on $m-m'$, the case where the currents are injected at points in other cells is straightforward: one has just to suitably translate the expressions of the components of the Green function. If for simplicity we set $m'=0$, then $m=-1$ refers to the cell on the left of cell $0$, i.e. cell $M-1$, and $m=+1$ refers
to the cell on the right of cell $0$, i.e. cell $1$. The expression of the components $G_{k,k'}(m-m')$ of the Green function so derived are provided in Section \ref{Green_function_2phases}, and the components of the associated transfer matrix, a $4M\times 4M$ matrix, are easily recovered by virtue of \eq{3.2}. For the space-time geometry considered here, all the eigenvalues of the transfer matrix lie on the unit circle so that the corresponding eigenvectors are all propagating modes. There are no modes that blow up with time.

Suppose one injects a unit current at one of the $4M$ points and wonders how such a current propagates through the space-time microstructure in \fig{check1} as time evolves. If we look at the expression of the components of the Green function, we see that $G(1,1,-1)=1$, $G(2,2,1)=1$, $G(3,3,-1)=1$, and $G(4,4,1)=1$, that is, the current injected at a certain point along a certain characteristic line will be flowing along the same characteristic line with the same intensity as time evolves. Note that, every time current is injected at a specific point, after one time period, it will reach 7 points: the point belonging to the same characteristic line will have the same current and the other 6 points, due to the conservation of current flux, will have currents whose sum is equal to zero. This implies also that, after the initial front has passed, the net current flowing across a unit cell is zero. Moreover, the current distribution along the characteristic lines is periodic: the intensity of the current along a certain characteristic line does not change in time, thus giving rise to a periodic solution, as shown in \fig{check1_inj1} which represents the evolution of the current flow with respect to time when we inject a unit current at point 1 of cell 50, that is, at the point labeled with 201. Furthermore, Video1 shows the evolution in time of the current flow and the electric potential for the field pattern of \fig{check1}: the electric potential, piecewise constant, has jumps in correspondence to the characteristic lines that are proportional to the modulus of the current flowing through them.
\begin{figure}[!ht]
	\centering
	\includegraphics[width=0.7\textwidth]{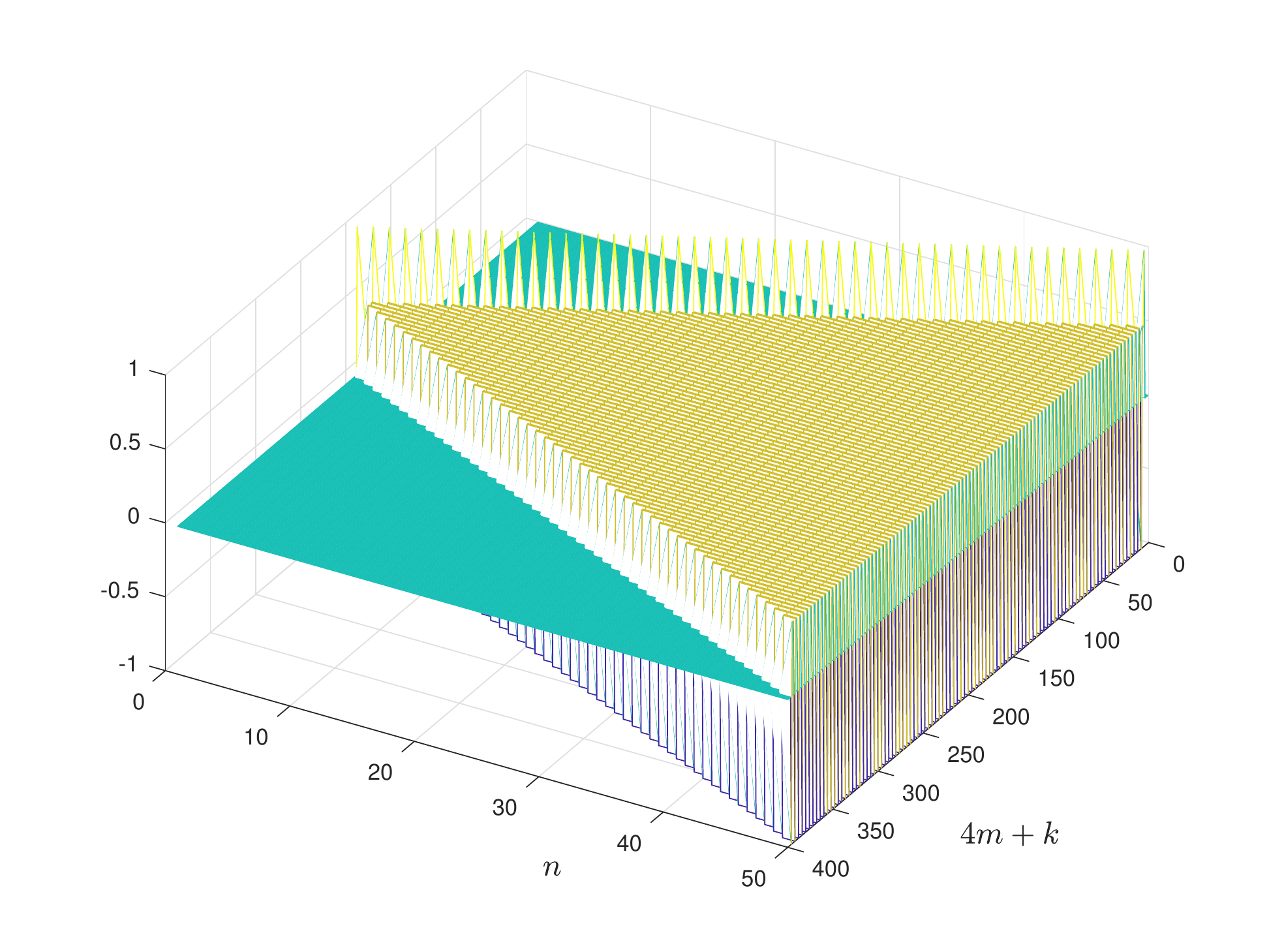}
	\caption{Evolution of the current flow in the space-time checkerboard of \fig{check1}, with $\Gg_1=1$ and $\Gg_2=3$, when 100 unit cells are considered (since there are 4 points to monitor in each unit cell, the total number of points is equal to 400). We inject a unit current at point 1 of cell 50, that is, at the point labeled with 201, and as time evolves, that is, for $t=\tau+nt_0$ with $n=0,1,\dots,50$, the current flows with the same intensity at the edge of a conic shape having vertex at the point of injection of current. Inside the conic shape the net current is equal to zero and the solution is periodic, that is, after each time period the modulus of the current at each point in space does not change. See also Video1.}
	\labfig{check1_inj1}     	
\end{figure}

Notice that, by suitably changing the slope of the space--time interfaces between the phases in the geometry of \fig{check1}, one could obtain an infinite number of different space-time checkerboards, each one providing the same field pattern. In fact, recalling that the global potential is piecewise constant due to the initial conditions \eq{1.7a} and that the jumps in potential occur along characteristic lines, one can change the slope of the space-time interfaces (within each area enclosed by the characteristics the potential will still be constant) so that the intersection points between the characteristic lines and the space-time interfaces do not change, thus obtaining an equivalent space-time microstructure. 
As an example, in \fig{alternative_check} we see that there are many space-time checkerboards composed by trapezoids, parallelograms, pentagons, and hexagons that are equivalent to the checkerboard illustrated in \fig{check1} for the given initial condition. 
\begin{figure}[ht!]
	\centering
	\begin{subfigure}{.45\linewidth}
		\includegraphics[width=\textwidth]{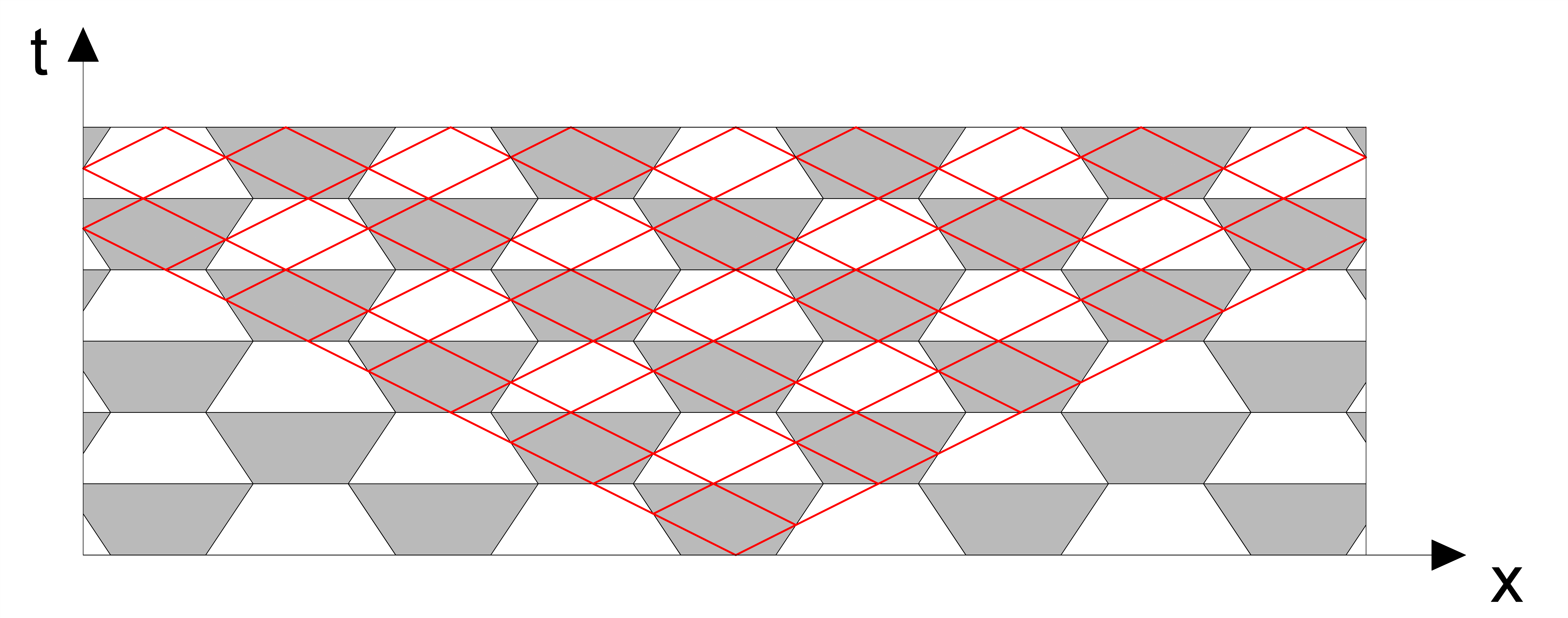}
	\end{subfigure}
	\hskip2em
	\begin{subfigure}{.45\linewidth}
		\includegraphics[width=\textwidth]{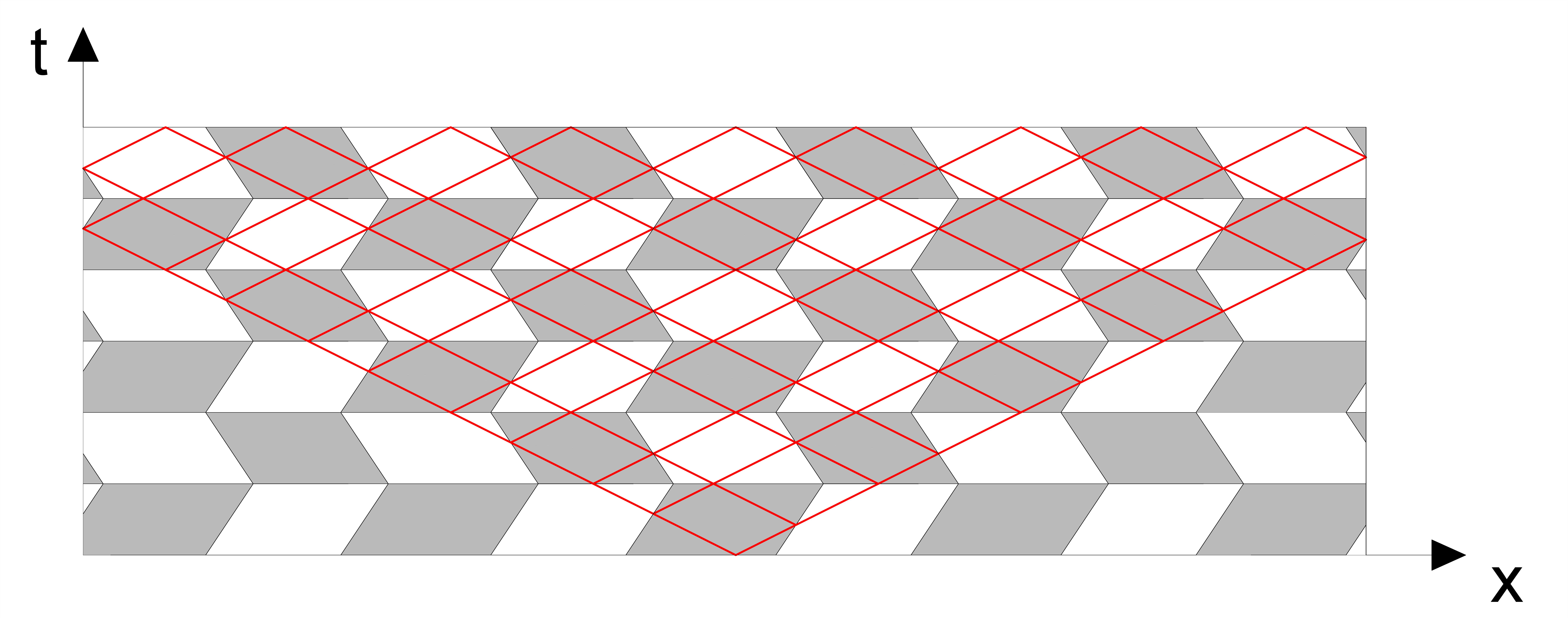}
	\end{subfigure}
	\begin{subfigure}{.45\linewidth}
		\includegraphics[width=\textwidth]{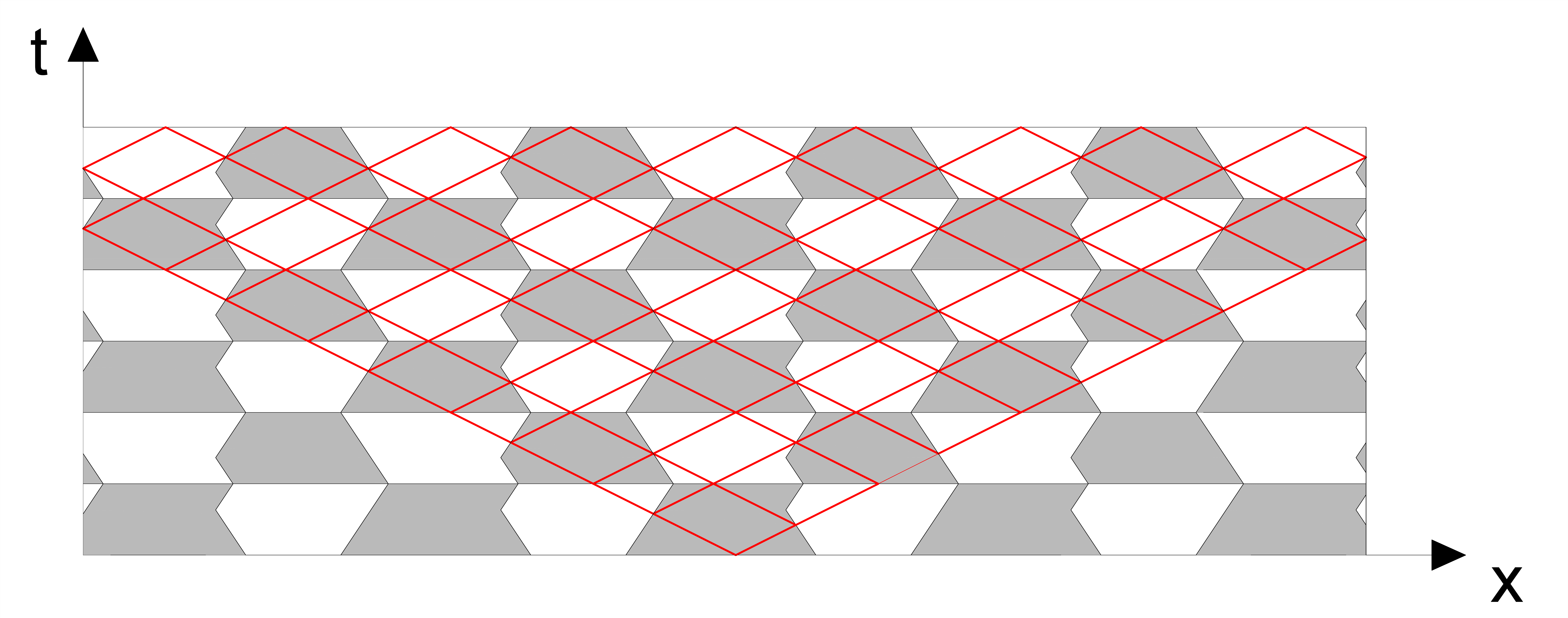}
	\end{subfigure}
	\hskip2em
	\begin{subfigure}{.45\linewidth}
		\includegraphics[width=\textwidth]{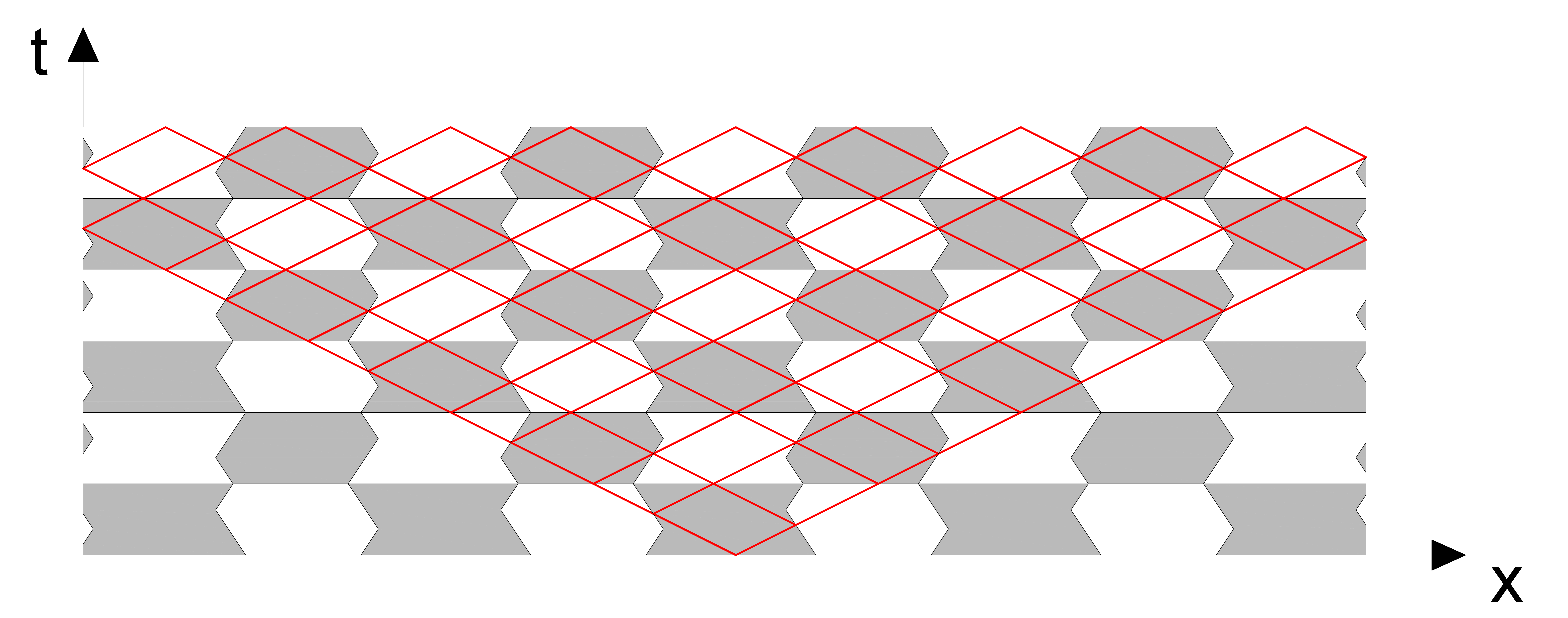}
	\end{subfigure}
	\caption{These are examples of space-time checkerboards giving rise to the same field pattern shown in \fig{check1} for the given initial conditions. Obviously, these new space-time geometries, realized by using trapezoids, parallelograms, pentagons and hexagons, have the same transfer matrix of the checkerboard of \fig{check1}. Notice that if we change the initial point where the disturbance was created then all of these space-time microstructures will have different responses. Note that these space--time microstructures,
with the exception of the top right one, do not have $\cal P\cal T$--symmetry, although the specific
field pattern considered here does have $\cal P\cal T$--symmetry.}
	\labfig{alternative_check}
\end{figure}

Now the question is, what happens when one considers a more-than-two-component space-time checkerboard with the constituents having the same wave speed? Are all the eigenvalues of the corresponding transfer matrix still on the unit circle? Let us consider then a three-component space-time checkerboard with phases having the same wave speed such as the one shown in \fig{3phase_check}.

\begin{figure}[!ht]
	\includegraphics[width=\textwidth]{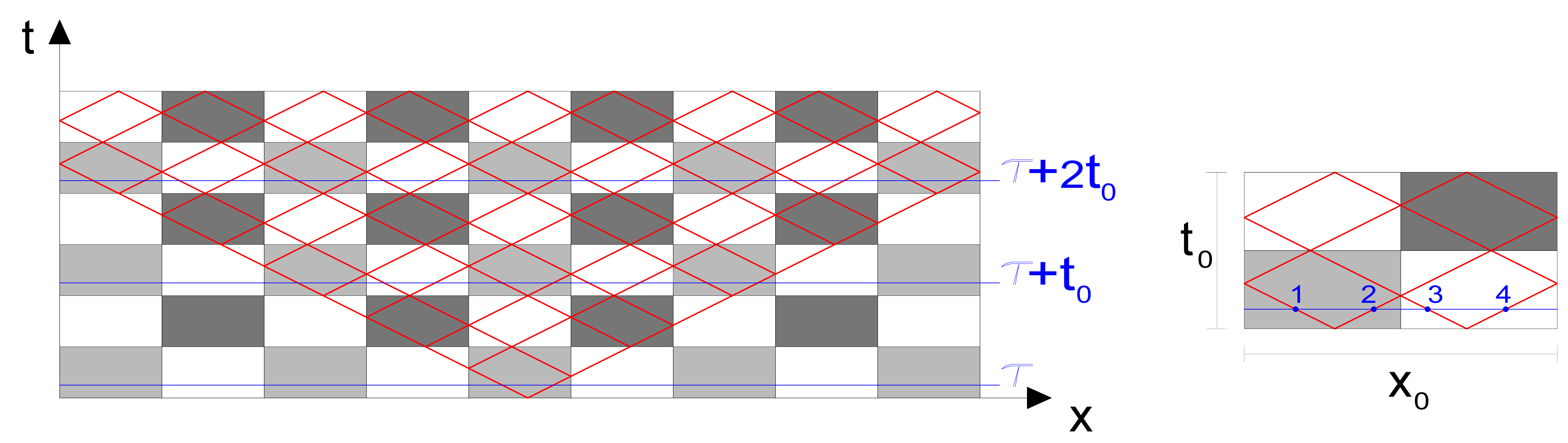}
	\caption{On the left is a three-phase space-time checkerboard with the three components having the same wave speed, $c_1=c_2=c_3=c$, giving rise to a field pattern (the red lines represent the characteristic lines). In particular, we denote phase 1 by the region colored in white, phase 2 by the one in light gray, and phase 3 by the one in dark gray. On the right, the unit cell of the dynamic network
is shown.}
	\labfig{3phase_check}     	
\end{figure}

The components of the associated Green function are given in Section \ref{Green_function_3phases}.
Also for this space-time checkerboard the transfer matrix has determinant equal to one and all its eigenvalues lie on the unit circle: all the eigenmodes are propagating modes. Furthermore, if one compares the expression of the components of the Green function for this microstructure with those for the two-phase checkerboard, it is clear that the same features leading to the evolution of current flow shown in \fig{check1_inj1} still hold. Therefore, if one injects current at one of the $4M$ points on the dynamic network taking place in the three-phase checkerboard, the result will be similar to the one depicted in \fig{check1_inj1}.

Clearly, even in this case, the field pattern shown in \fig{3phase_check} can be obtained by an infinite number of space-time microstructures equivalent to the three-component checkerboard illustrated in \fig{3phase_check}.

Finally, let us consider a four-component space-time checkerboard with the four phases having the same wave speed, i.e., $c_1=c_2=c_3=c_4=c$, like the microstructure illustrated in \fig{4phase_check}.

\begin{figure}[!ht]
	\includegraphics[width=\textwidth]{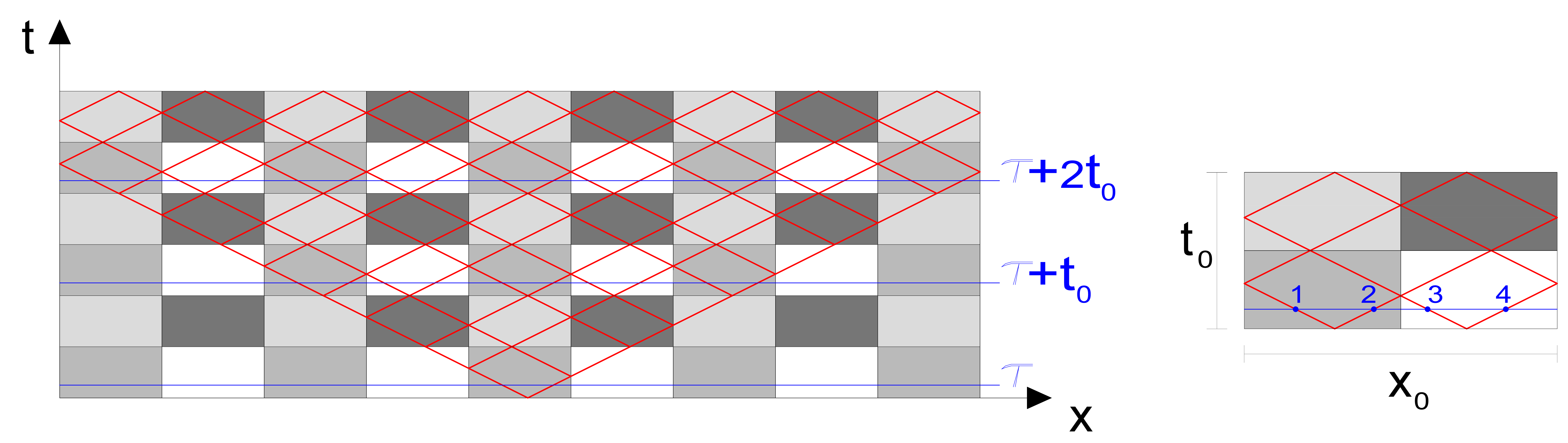}
	\caption{On the left is a four-phase space-time checkerboard with the four components having the same wave speed, $c_1=c_2=c_3=c_4=c$, giving rise to a field pattern (the red lines represent the characteristic lines). In particular, we denote phase 1 by the region colored in white, phase 2 by the one in light gray, phase 3 by the one in dark gray, and phase 4 by the one in very light gray. On the right, the unit cell of the dynamic network is shown.}
	\labfig{4phase_check}     	
\end{figure}

The expression of the Green function for this case, whose components are provided in Section \ref{Green_function_4phases}, is significantly different from the one corresponding to the two-phase and three-phase checkerboard geometries and this yields very interesting results. First of all, the distribution of the eigenvalues of the transfer matrix in the complex plane depends on the ratio between the coefficients $\Gg_1$, $\Gg_2$, $\Gg_3$ and $\Gg_4$. In particular, by numerical exploration, we observe that among the 24 combinations in which one can order these 4 parameters, only 12 seem to lead to the case of the so-called ``unbroken $\CP\CT$--symmetry'' (e.g., \cite{Bender:2007:MSN}) for which all the eigenvalues have unit modulus. In particular, the numerical evidence points to this happening when and only when the parameters are chosen in the following orderings:
\beq
\begin{aligned}
	&\Gg_1\geq \Gg_3\geq\Gg_2\geq\Gg_4,\qquad\Gg_1\geq \Gg_3\geq\Gg_4\geq\Gg_2,\qquad\Gg_1\geq \Gg_4\geq\Gg_3\geq\Gg_2,\\
	&\Gg_2\geq \Gg_3\geq\Gg_4\geq\Gg_1,\qquad\Gg_2\geq \Gg_4\geq\Gg_3\geq\Gg_1,\qquad\Gg_2\geq \Gg_4\geq\Gg_1\geq\Gg_3,\\
	&\Gg_3\geq \Gg_1\geq\Gg_2\geq\Gg_4,\qquad\Gg_3\geq \Gg_1\geq\Gg_4\geq\Gg_2,\qquad\Gg_3\geq \Gg_2\geq\Gg_1\geq\Gg_4,\\
	&\Gg_4\geq \Gg_1\geq\Gg_2\geq\Gg_3,\qquad\Gg_4\geq \Gg_2\geq\Gg_1\geq\Gg_3,\qquad\Gg_4\geq \Gg_2\geq\Gg_3\geq\Gg_1.\\
\end{aligned}
\eeq{combin}
Interestingly enough, in such a case, the distribution of the eigenvalues on the unit circle has a gap, as shown, for instance, on the left of \fig{check4_prop}, illustrating the eigenvalues of the transfer matrix for the case in which $\Gg_4\geq \Gg_1\geq\Gg_2\geq\Gg_3$. The corresponding eigenvectors are all propagating modes, as shown in \fig{check4_waves} representing the evolution of the current flow in time when two specific eigenvectors are applied as initial distribution of currents. Furthermore, if we inject a unit current at a specific point in the discrete network, as on the right of \fig{check4_prop}, we obtain a solution quite different from the one related to the two-phase space-time checkerboard case (see \fig{check1_inj1}), according to the differences in the expression of the associated Green functions: the current disperses along the characteristic lines and the solution inside the conic shape is not periodic anymore.

\begin{figure}[ht!]
	\centering
	\begin{subfigure}{.45\linewidth}
		\includegraphics[width=\textwidth]{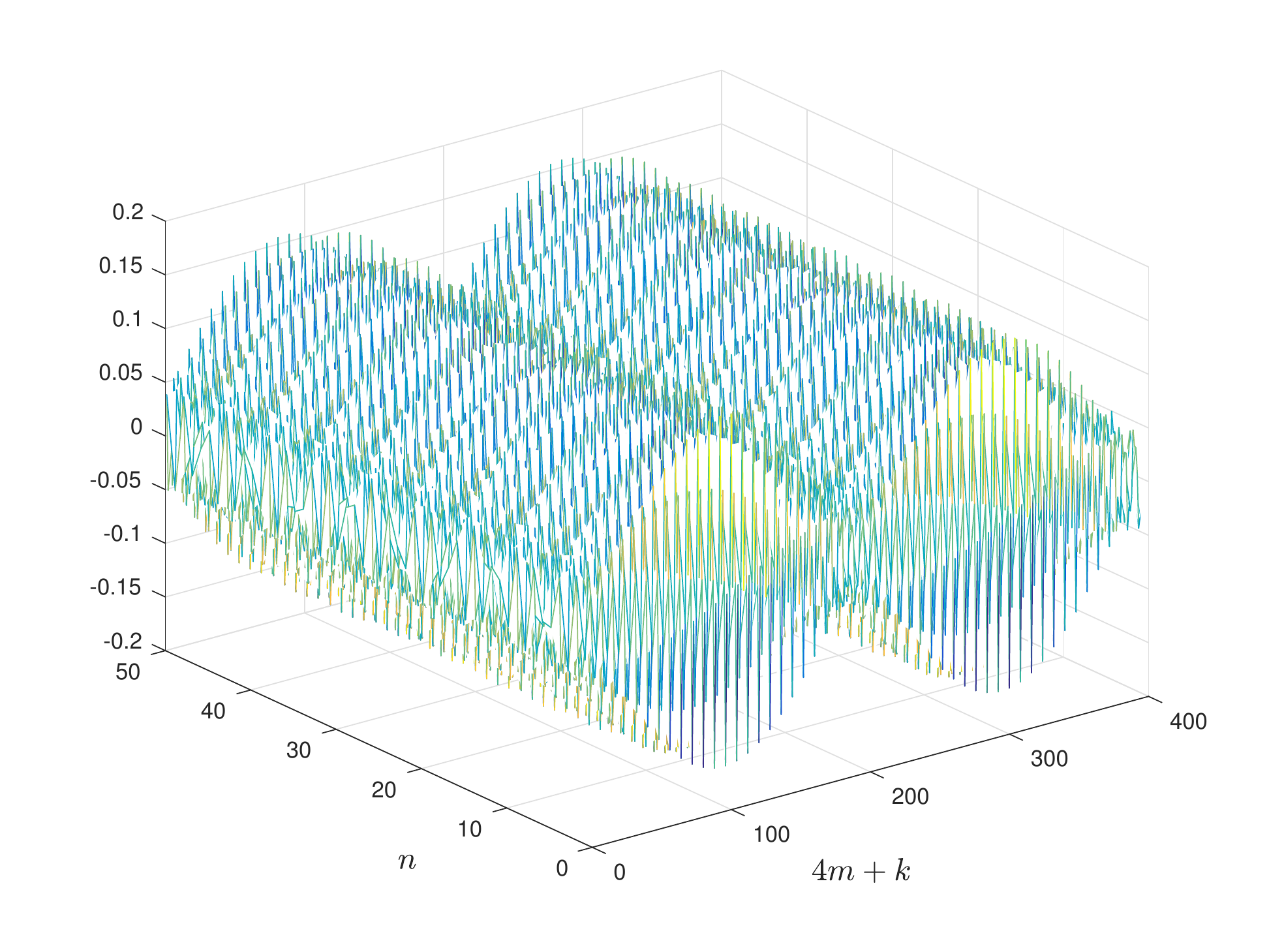}
	\end{subfigure}
	\hskip2em
	\begin{subfigure}{.45\linewidth}
		\includegraphics[width=\textwidth]{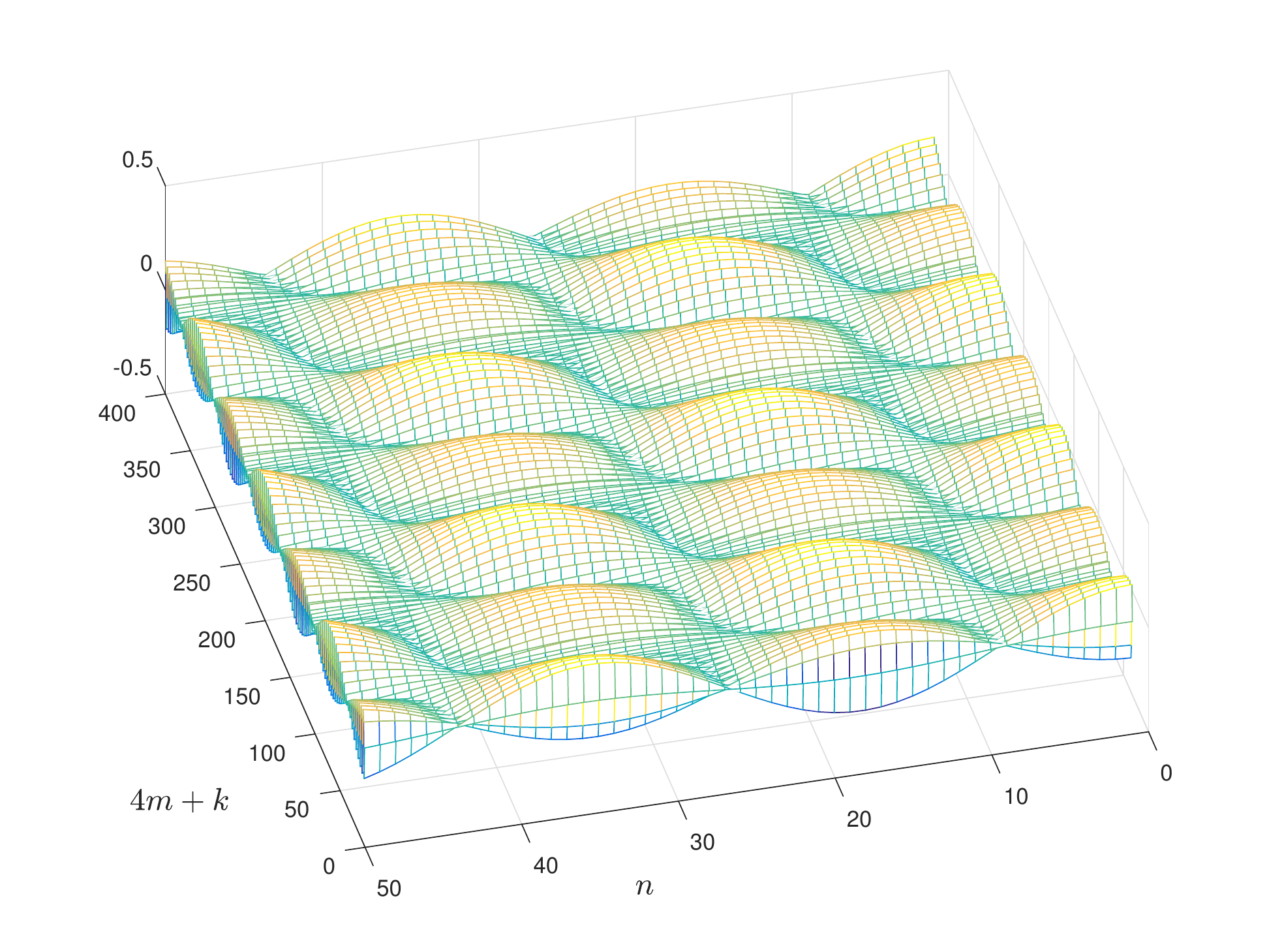}
	\end{subfigure}
	\caption{Evolution of the current flow in 100 cells for $t=\tau+nt_0$ with $n=0,1,...,100$, in the case when $\Gg_1=5$, $\Gg_2=2$, $\Gg_3=1$ and $\Gg_4=11$ and, at $t=\tau$ ($n=0$), we inject a distribution of currents equal to that given by the sum of one of the couples of two conjugate eigenvectors corresponding to the eigenvalues $-0.8553 \pm 0.5181i$ (on the left) and to the eigenvalues $0.9836 \pm 0.1806i$ (on the right). In particular, the solutions are periodic in time and space, but the periodicity is not that of the unit cell.}
	\labfig{check4_waves}
\end{figure}

\begin{figure}[ht!]
	\centering
	\begin{subfigure}{.45\linewidth}
		\includegraphics[width=\textwidth]{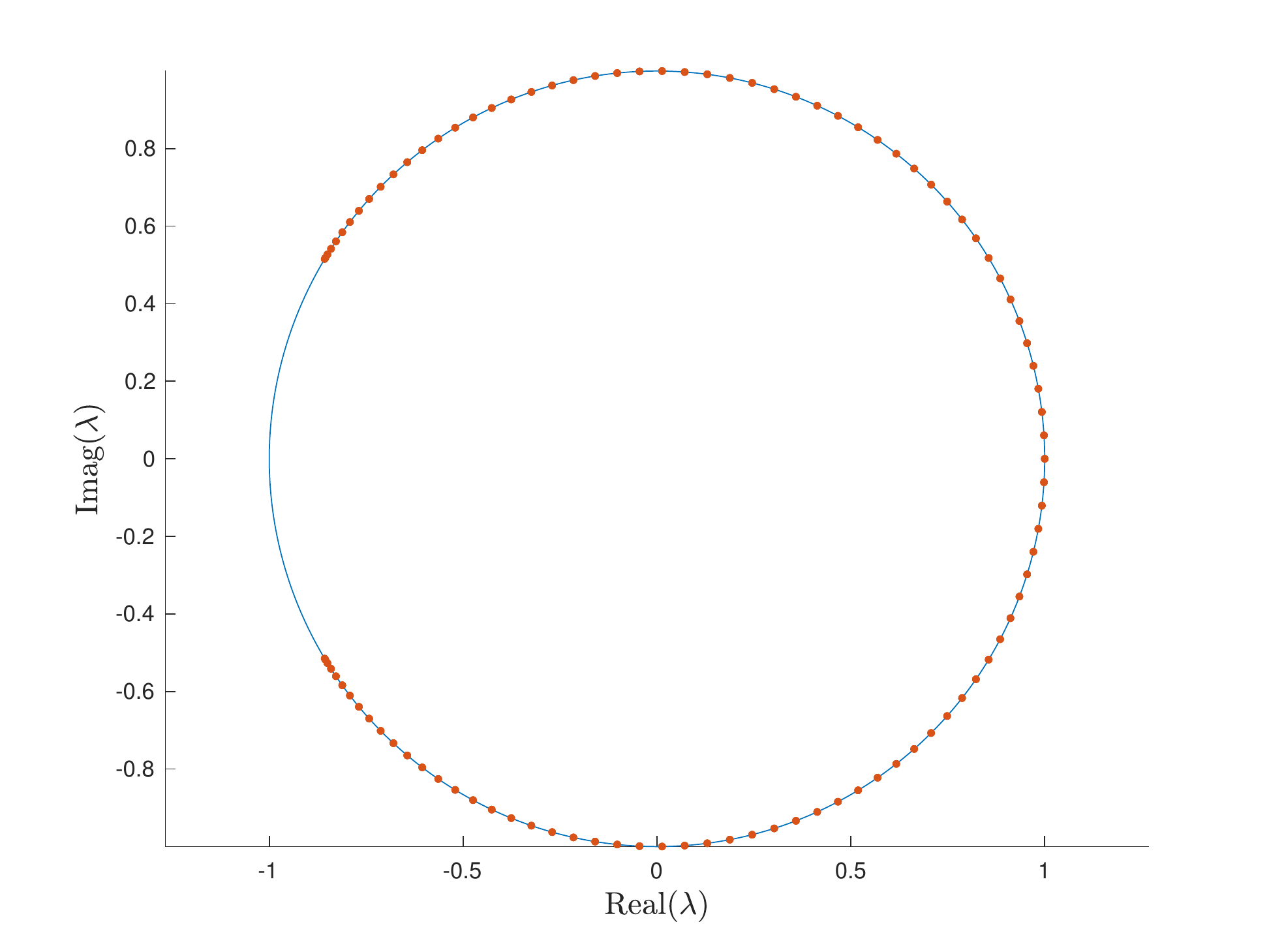}
	\end{subfigure}
	\hskip2em
	\begin{subfigure}{.45\linewidth}
		\includegraphics[width=\textwidth]{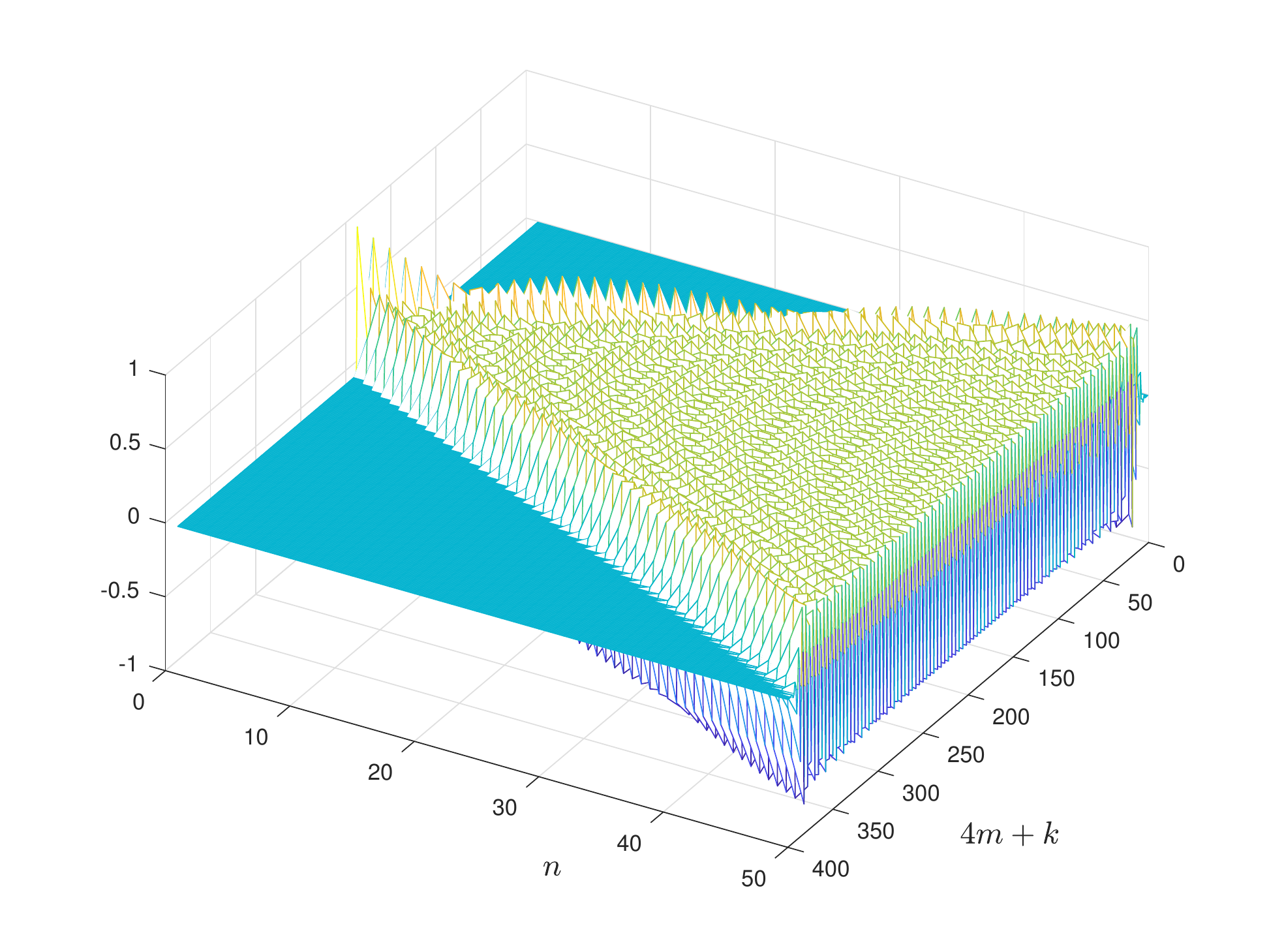}
	\end{subfigure}
	\caption{On the left, we see that the eigenvalues of the transfer matrix for the four-component space-time checkerboard with $\Gg_1=5$, $\Gg_2=2$, $\Gg_3=1$ and $\Gg_4=11$ all have unit modulus and their distribution on the unit circle leaves a gap. In this case we consider 100 unit cells and, since there are 4 relevant points in each cell, the transfer matrix turns out to be a $400\times 400$ matrix with 400 eigenvalues. On the right is shown the corresponding evolution in time of the current flow when a unit current is injected at point 1 of cell 50 (the point labeled with 201).}
	\labfig{check4_prop}
\end{figure}

When the parameters $\Gg_1$, $\Gg_2$, $\Gg_3$ and $\Gg_4$ do not fulfill any of the relations presented in \eq{combin},  the 
numerical results indicate that one always has ``broken $\CP\CT$--symmetry'' (e.g., \cite{Bender:2007:MSN}), the eigenvalues of the transfer matrix can have modulus equal, greater or smaller than one, thus giving rise, respectively, to propagating modes, modes that blow up exponentially with time and modes that decrease exponentially with time. In particular, the eigenvalues that do not lie on the unit circle are all real, as shown on the left of \fig{check4_blow}. When one injects current at one point of the discrete network, the solution blows up exponentially with time, as depicted on the right of \fig{check4_blow} where the growth in time of the current distribution is captured using a logarithmic scale for the amplitude. 
\begin{figure}[ht!]
	\centering
	\begin{subfigure}{.45\linewidth}
		\includegraphics[width=\textwidth]{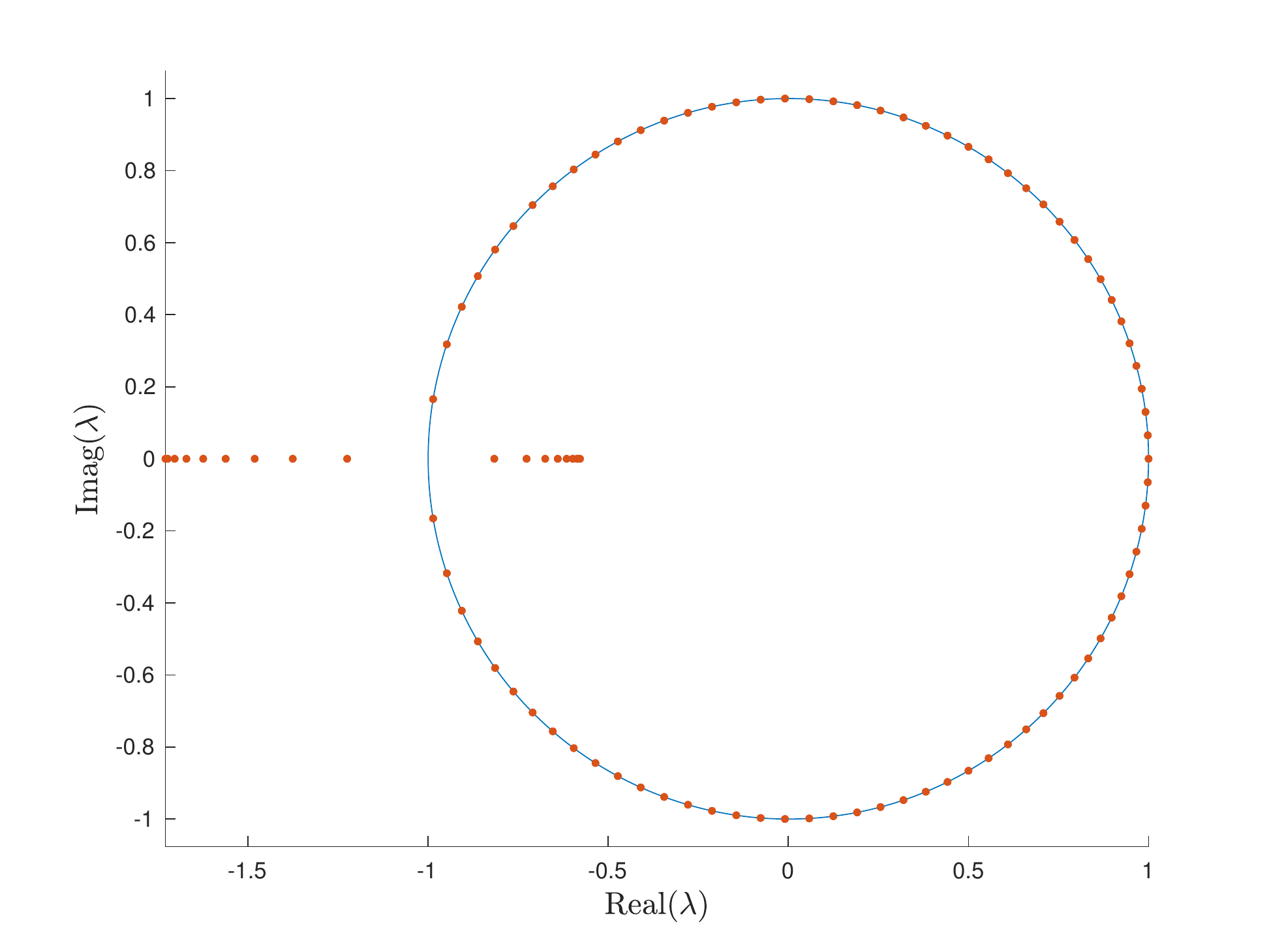}
	\end{subfigure}
	\hskip2em
	\begin{subfigure}{.45\linewidth}
		\includegraphics[width=\textwidth]{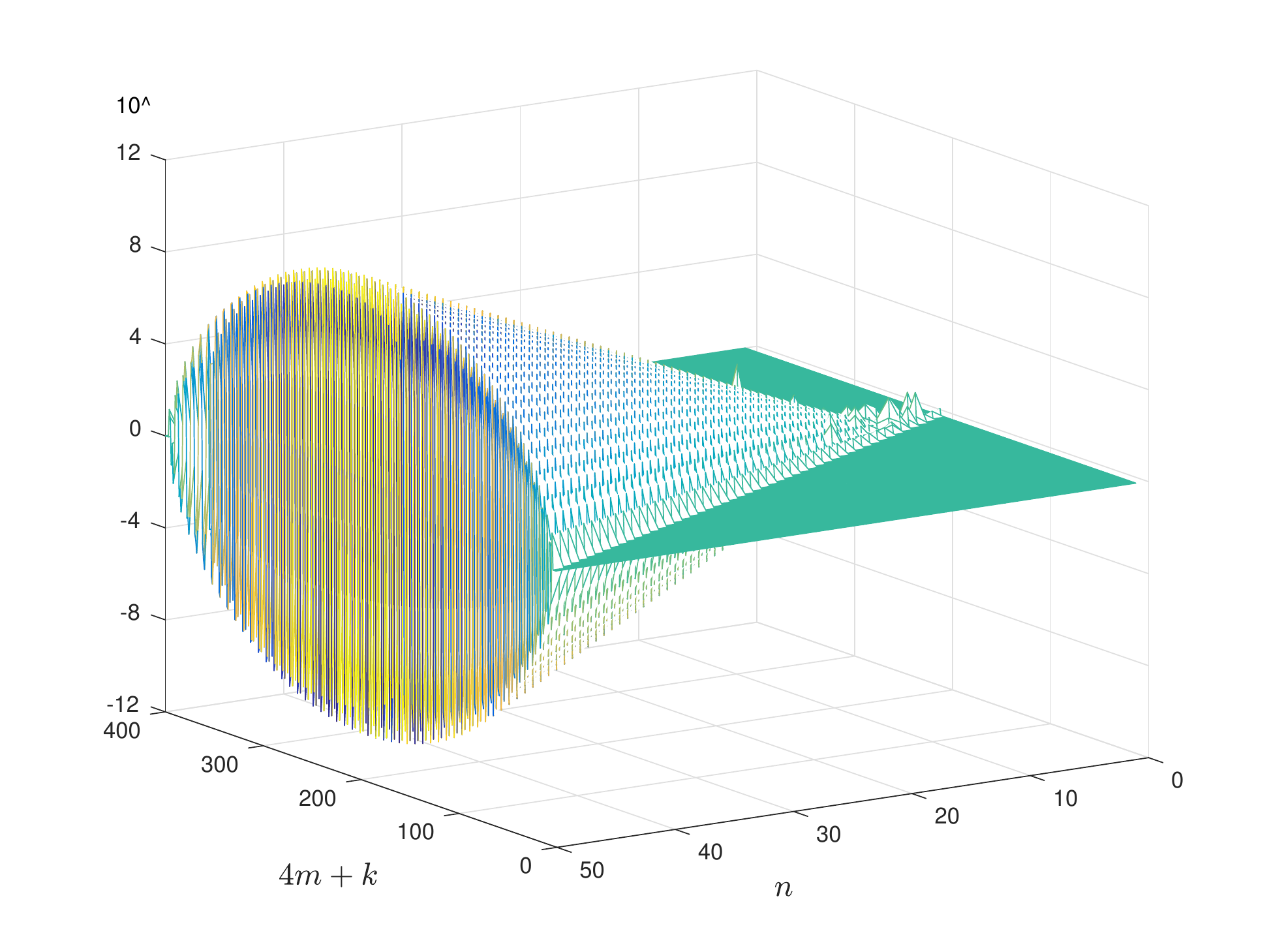}
	\end{subfigure}
	\caption{On the left, we see that the eigenvalues of the transfer matrix for the four-component space-time checkerboard with $\Gg_1=5$, $\Gg_2=1$, $\Gg_3=2$ and $\Gg_4=11$ either are real or lie on the unit circle. Here we consider 100 unit cells so, since there are 4 relevant points in each unit cell, the transfer matrix turns out to be a $400\times 400$ matrix, for a total of 400 eigenvalues. On the right
is shown the corresponding evolution in time of the current flow when a unit current is initially injected at point 1 of cell 50 (the point labeled with 201) with a logarithmic vertical 
scale: the solution blows up exponentially with time.}
	\labfig{check4_blow}
\end{figure}

\section{Three-component space-time checkerboards with phases having wave speeds in a certain ratio}\label{Check_spec_ratio}

Another example of space-time geometries giving rise to field patterns is the case of three-component space-time checkerboards in which the ratio between the wave speeds of the three phases is constant, say, $c_2/c_1=c_1/c_3=r$, with $r$ an integer. \fig{3phase_ratio2} and \fig{3phase_ratio3} depict the case in which $r=2$ and $r=3$, respectively (if phase 1 is the one colored in white, phase 2 the one in light gray and phase 3 the one in dark gray).
\begin{figure}[!ht]
	\includegraphics[width=\textwidth]{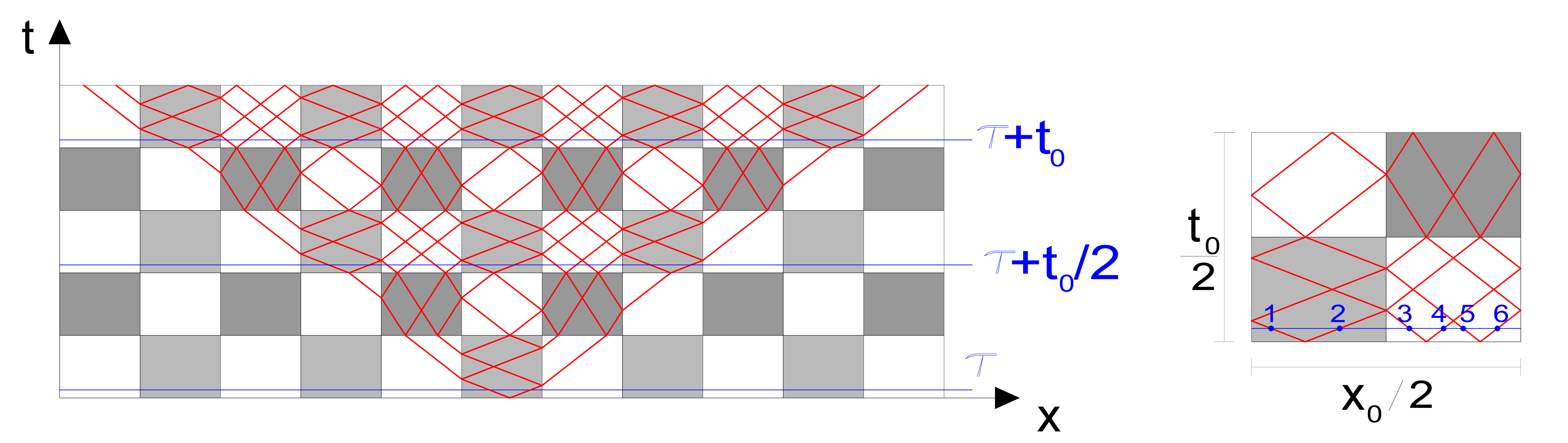}
	\caption{On the left is shown another example of a space-time geometry giving rise to a field pattern: it consists of a three-phase space-time checkerboard with the wave speed of the three components chosen such that $c_2/c_1=c_1/c_3=2$, with phase 1 the one colored in white, phase 2 the one in light gray, and phase 3 the one in dark gray. On the right, due to the translational symmetry in space and time of the dynamic network we represent only half of the unit cell of periodicity.}
	\labfig{3phase_ratio2}     	
\end{figure} 
\begin{figure}[!ht]
	\includegraphics[width=\textwidth]{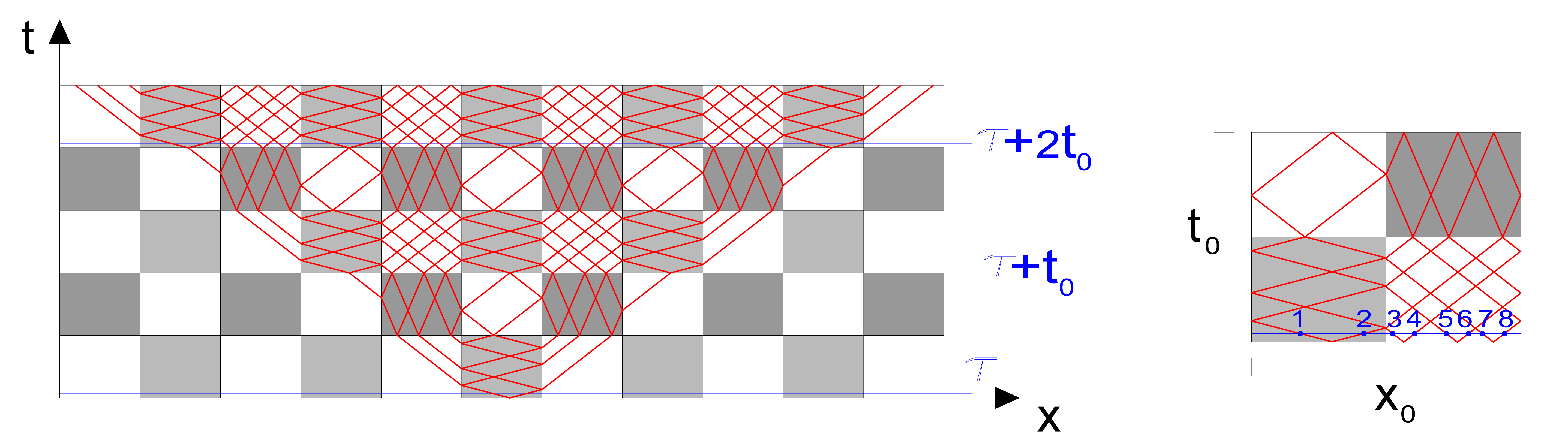}
	\caption{On the left is another example of space-time geometry giving rise to a field pattern: it consists of a three-phase space-time checkerboard with the wave speeds of the three components chosen such that $c_2/c_1=c_1/c_3=3$, with phase 1 the one colored in white, phase 2 the one in light gray, and phase 3 the one in dark gray. On the right, we show the unit cell of periodicity.}
	\labfig{3phase_ratio3}     	
\end{figure}

Due to the complicated expression of the components of the Green function associated with the two space-time microstructures above, we do not include them but we recall that, as usual, to recover the components of the Green function one has just to inject a unit current at each of the relevant points of the unit cell (6 for the space-time geometry depicted in \fig{3phase_ratio2} and 8 for the one in \fig{3phase_ratio3}) and calculate how the current flows along the characteristic lines to determine the current distribution after one time period. In particular, for the microstructure in \fig{3phase_ratio2} the current injected at a point reaches 11 points after one time period, whereas for the microstructure in \fig{3phase_ratio3} it reaches 15 points. 

For the space-time checkerboard of \fig{3phase_ratio2}, the distribution of the eigenvalues of the transfer matrix depends on the choice of the three parameters $\Gg_1$, $\Gg_2$ and $\Gg_3$. In particular, by numerical exploration we found that, if $\Gg_1\leq \Gg_2\leq \Gg_3$, then the scenario is that shown in \fig{ratio2_case123}; if $\Gg_3\leq \Gg_1$ and $\Gg_3\leq \Gg_2$, it is the one depicted in \fig{ratio2_case312}; otherwise, the situation is the one represented in \fig{ratio2_case231}.
\begin{figure}[ht!]
	\centering
	\begin{subfigure}{.45\linewidth}
		\includegraphics[width=\textwidth]{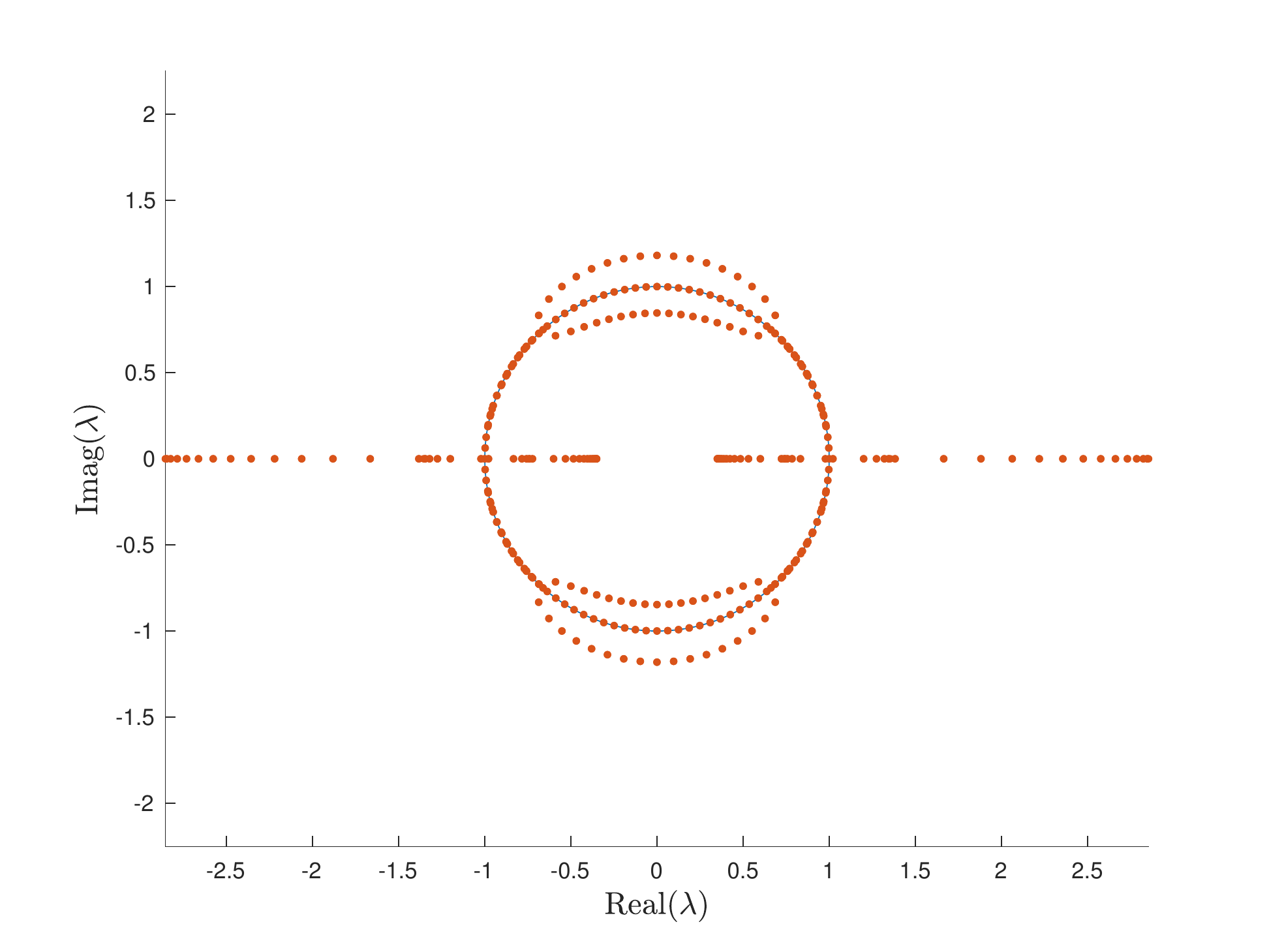}
	\end{subfigure}
	\hskip2em
	\begin{subfigure}{.45\linewidth}
		\includegraphics[width=\textwidth]{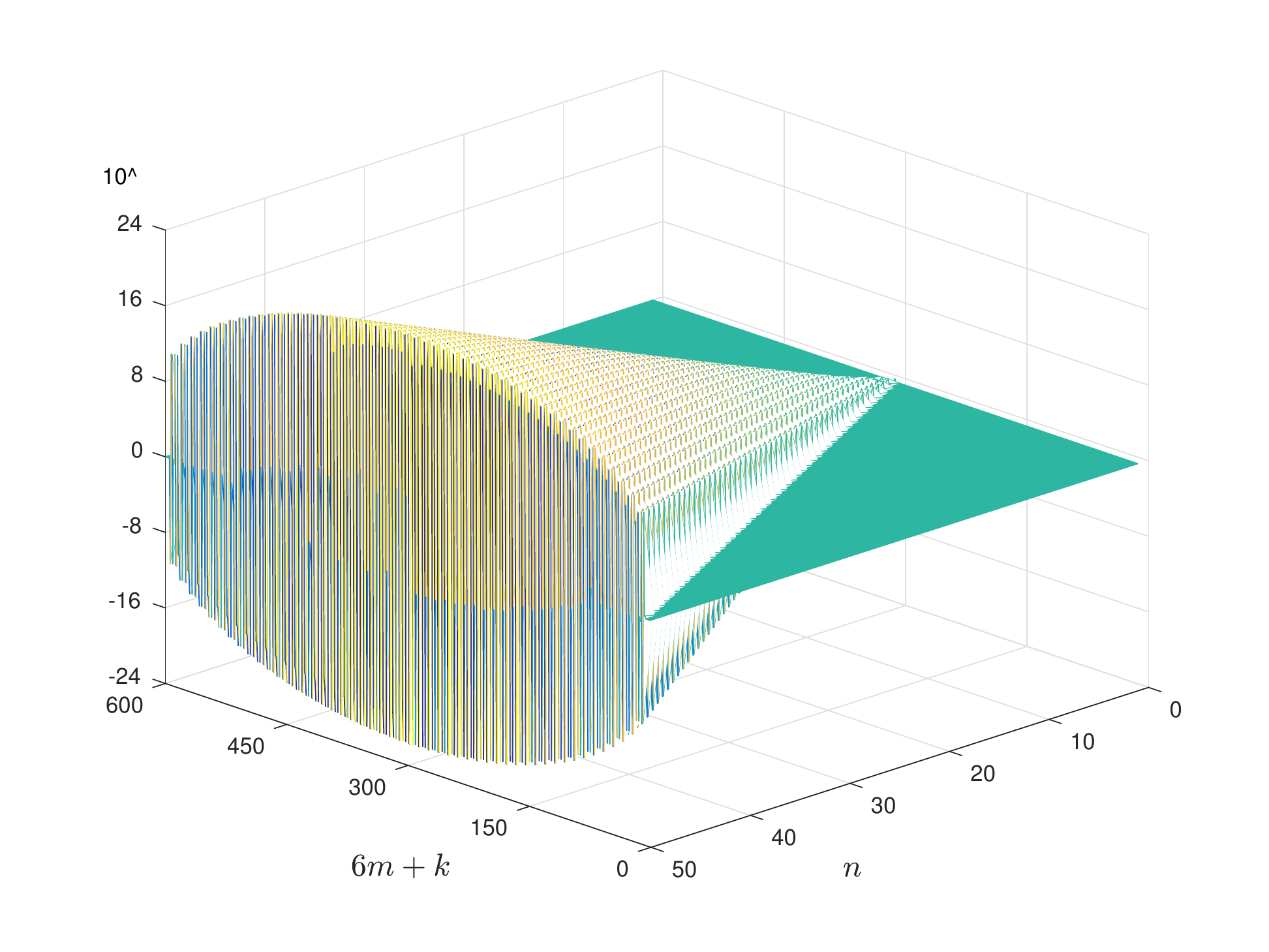}
	\end{subfigure}
	\caption{On the left is the distribution of the eigenvalues of the transfer matrix for the three-component space-time checkerboard of \fig{3phase_ratio2} when $\Gg_1=1$, $\Gg_2=7$, and $\Gg_3=13$. In this case we consider 100 unit cells and, since there are 6 relevant points in each cell, the transfer matrix turns out to be a $600\times 600$ matrix with 600 eigenvalues. On the right, we see the corresponding evolution in time of the current flow when a unit current is initially injected at point 1 of cell 50 (the point labeled with 301) with a logarithmic vertical scale: the solution blows up exponentially with time.}
	\labfig{ratio2_case123}
\end{figure}
\begin{figure}[ht!]
	\centering
	\begin{subfigure}{.45\linewidth}
		\includegraphics[width=\textwidth]{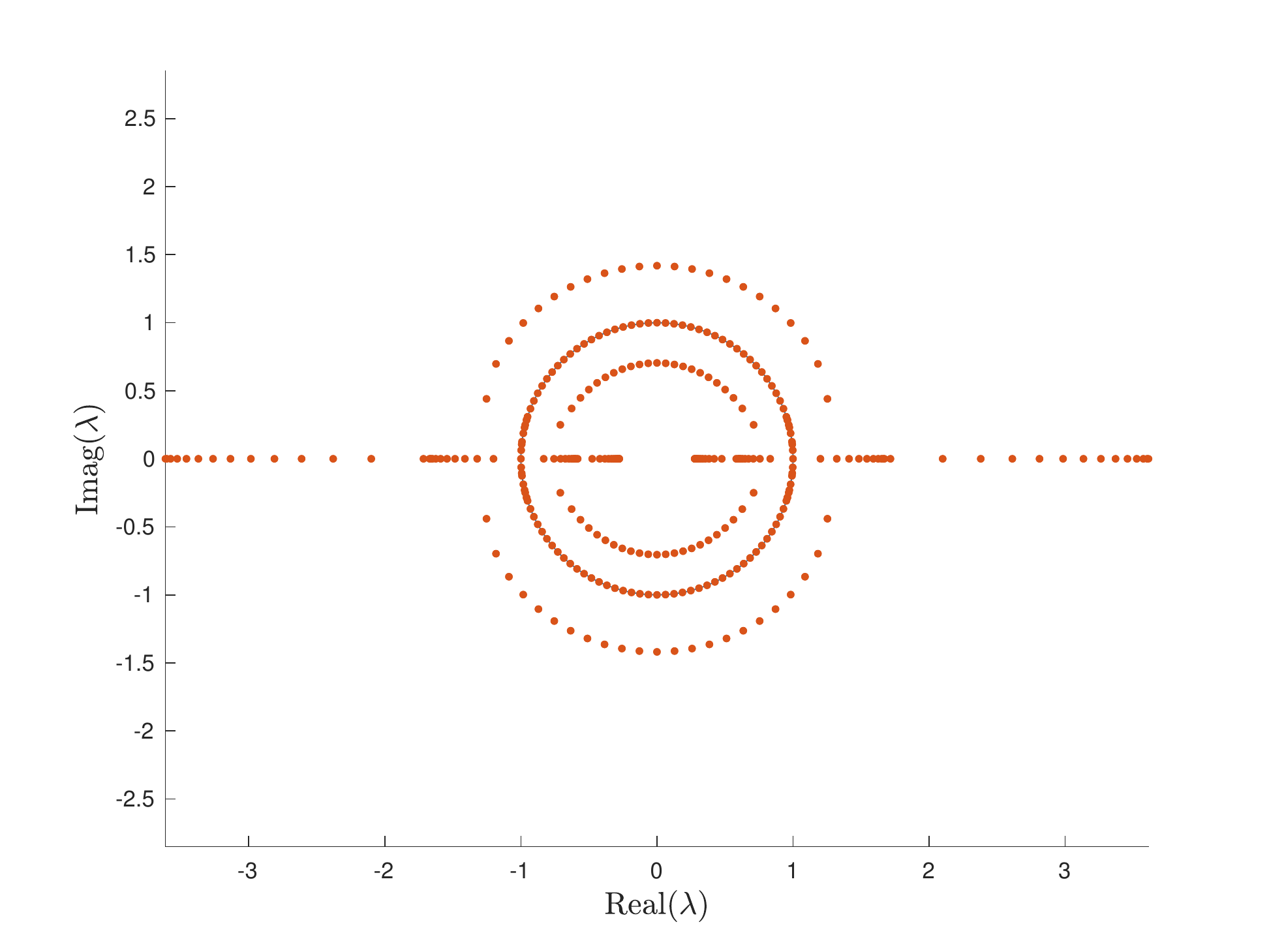}
	\end{subfigure}
	\hskip2em
	\begin{subfigure}{.45\linewidth}
		\includegraphics[width=\textwidth]{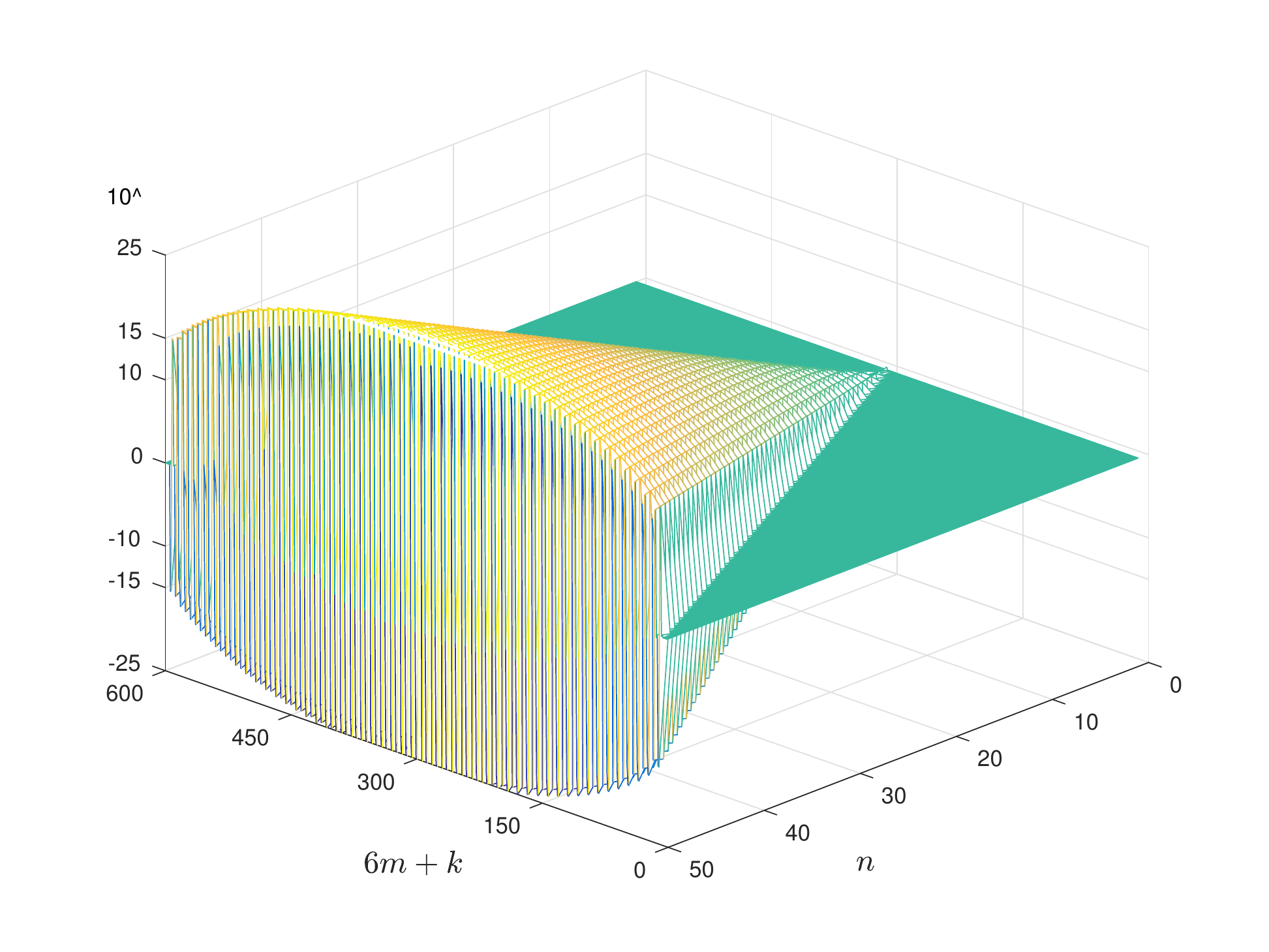}
	\end{subfigure}
	\caption{Similar to \fig{ratio2_case123}, but with parameters $\Gg_1=7$, $\Gg_2=13$, and $\Gg_3=1$.}
	\labfig{ratio2_case312}
\end{figure}
\begin{figure}[ht!]
	\centering
	\begin{subfigure}{.45\linewidth}
		\includegraphics[width=\textwidth]{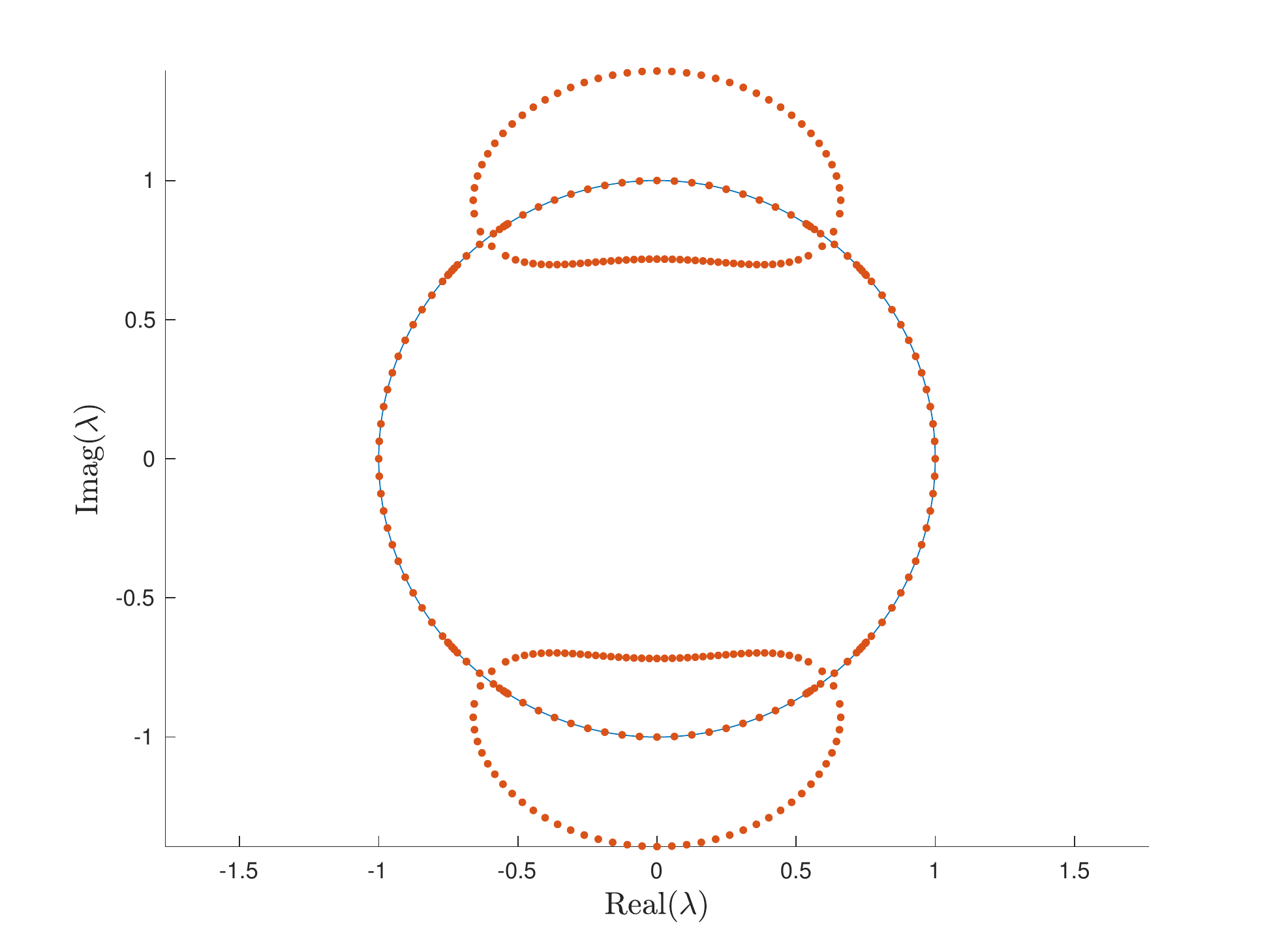}
	\end{subfigure}
	\hskip2em
	\begin{subfigure}{.45\linewidth}
		\includegraphics[width=\textwidth]{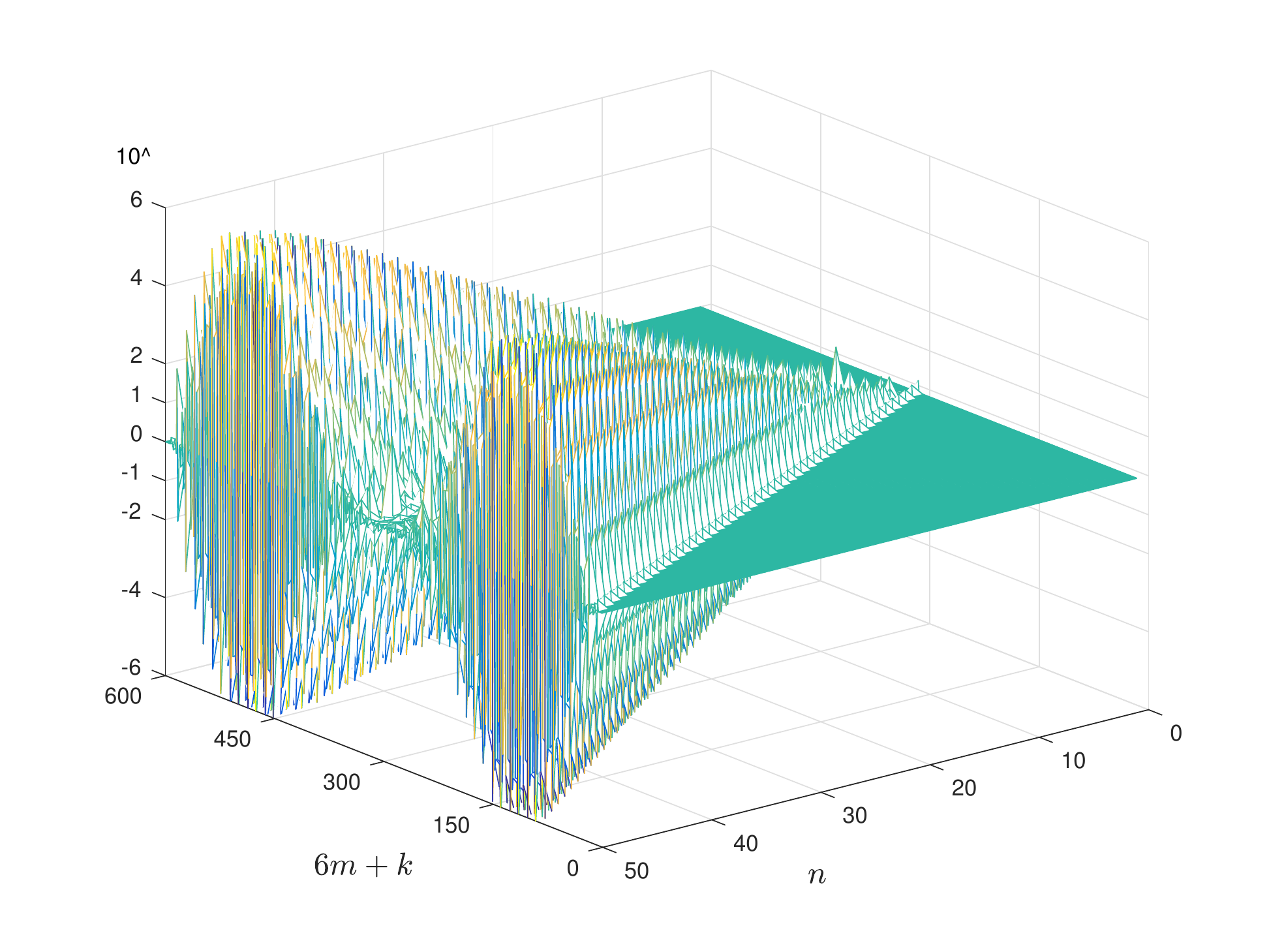}
	\end{subfigure}
	\caption{Similar to \fig{ratio2_case123} and \fig{ratio2_case312}, but with parameters $\Gg_1=13$, $\Gg_2=1$, and $\Gg_3=7$.}
	\labfig{ratio2_case231}
\end{figure}
In each scenario, the eigenvalues do not all lie on the unit circle: the $\CP\CT$--symmetry condition is broken. Therefore, the modes associated with the space-time checkerboard of \fig{3phase_ratio2} could be propagating modes, growing modes or decreasing modes.

Very different results are obtained for the space-time geometry of \fig{3phase_ratio3}. Numerically we see that two possible scenarios can occur: if $\Gg_3\leq \Gg_1$ and $\Gg_3\leq \Gg_2$ or if $\Gg_1\leq \Gg_2\leq \Gg_3$, then we are in the case of ``broken $\CP\CT$--symmetry'' and the eigenvalues are distributed as in \fig{ratio3_case321}; otherwise, the eigenvalues are all on the unit circle (``unbroken $\CP\CT$--symmetry'') and the corresponding eigenmodes are all propagating modes. For example, \fig{check3pha_waves} represents the time evolution of the current distribution when we apply two specific eigenvectors as initial distributions of currents. If one considers this last case and injects a unit current at a specific point at $t=0$, the evolution of the distribution of the currents within the discrete network is shown in \fig{check3phases_inj1}: there is no blow up. 

\begin{figure}[ht!]
	\centering
	\begin{subfigure}{.45\linewidth}
		\includegraphics[width=\textwidth]{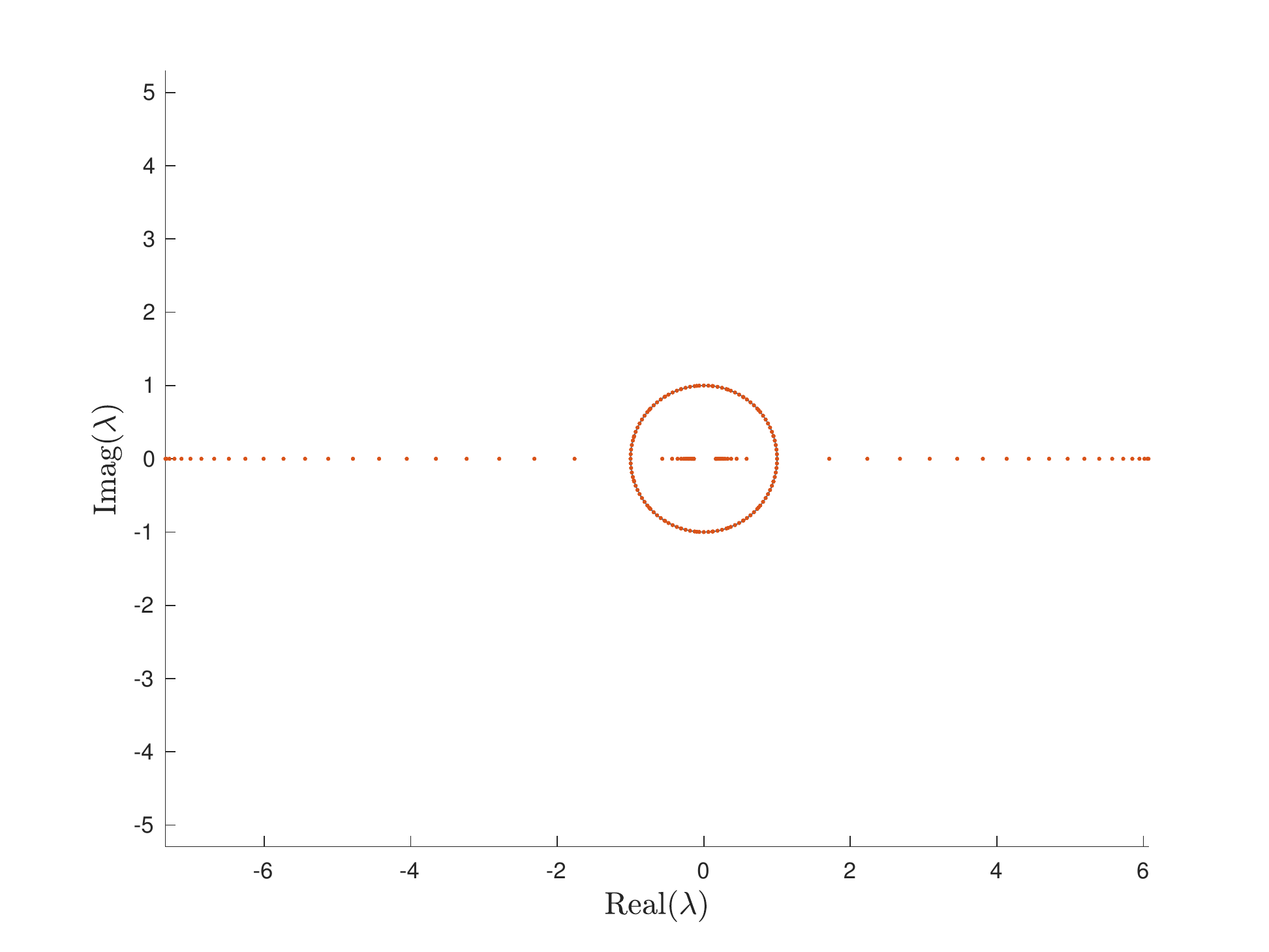}
	\end{subfigure}
	\hskip2em
	\begin{subfigure}{.45\linewidth}
		\includegraphics[width=\textwidth]{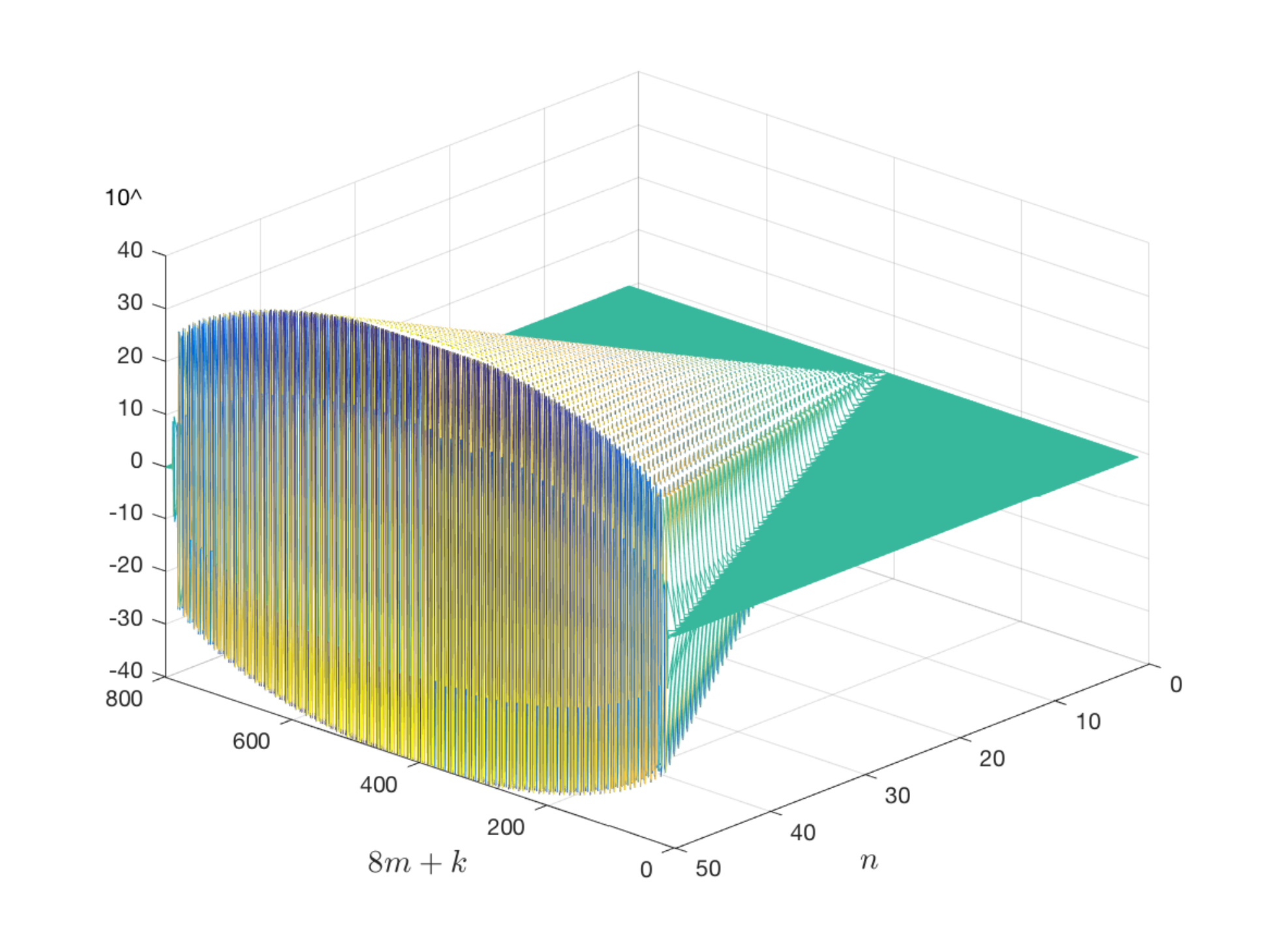}
	\end{subfigure}
	\caption{On the left is the distribution of the eigenvalues of the transfer matrix for the three-component space-time checkerboard of \fig{3phase_ratio3} when $\Gg_1=13$, $\Gg_2=7$, and $\Gg_3=1$. In this case we consider 100 unit cells and, since there are 8 relevant points in each cell, the transfer matrix turns out to be a $800\times 800$ matrix with 800 eigenvalues. On the right, we see the corresponding evolution in time of the current flow when a unit current is initially injected at point 1 of cell 50 (the point labeled with 401) with a logarithmic vertical scale: the solution blows up exponentially with time.}
	\labfig{ratio3_case321}
\end{figure}

\begin{figure}[ht!]
	\centering
	\begin{subfigure}{.45\linewidth}
		\includegraphics[width=\textwidth]{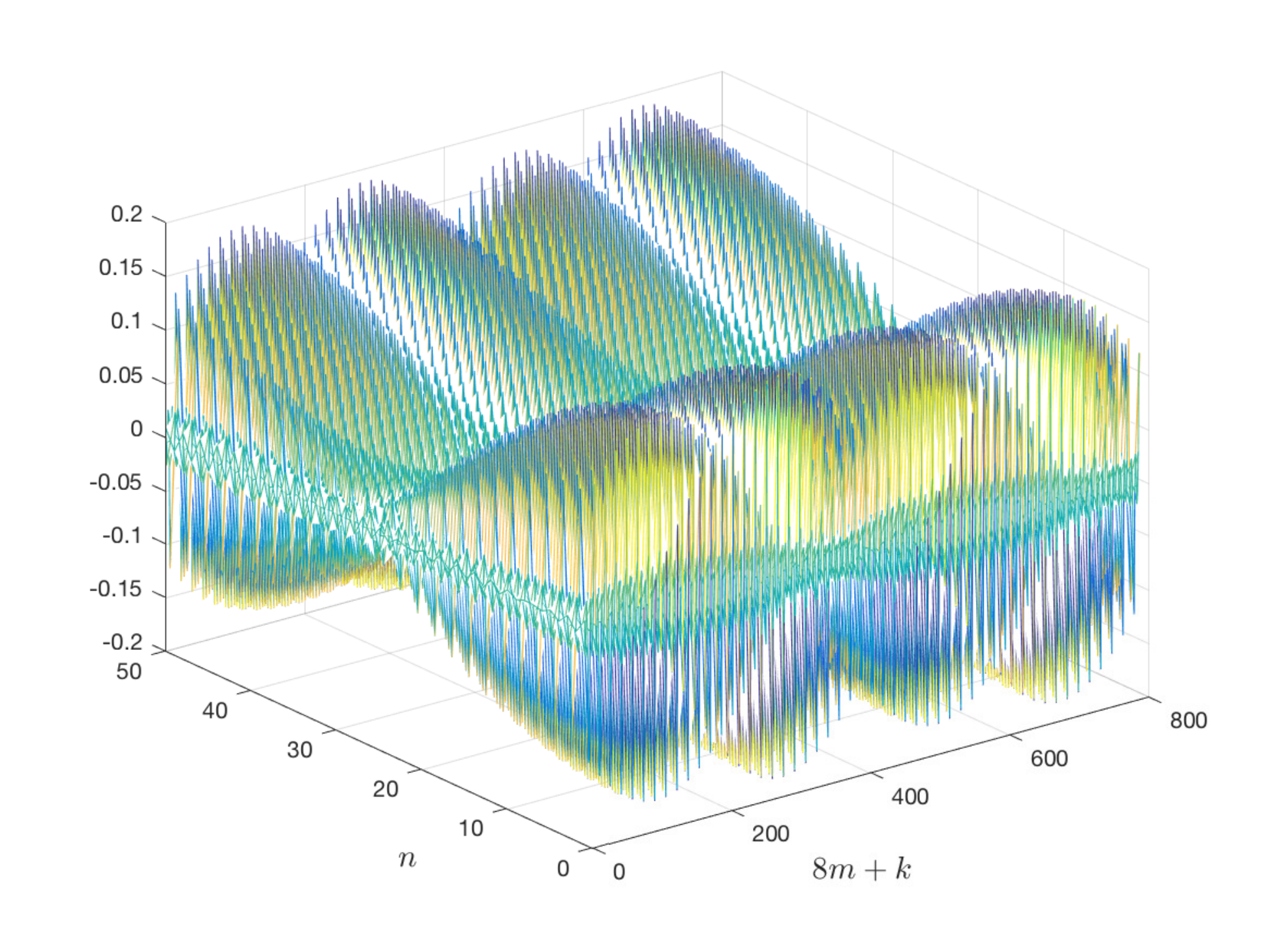}
	\end{subfigure}
	\hskip2em
	\begin{subfigure}{.45\linewidth}
		\includegraphics[width=\textwidth]{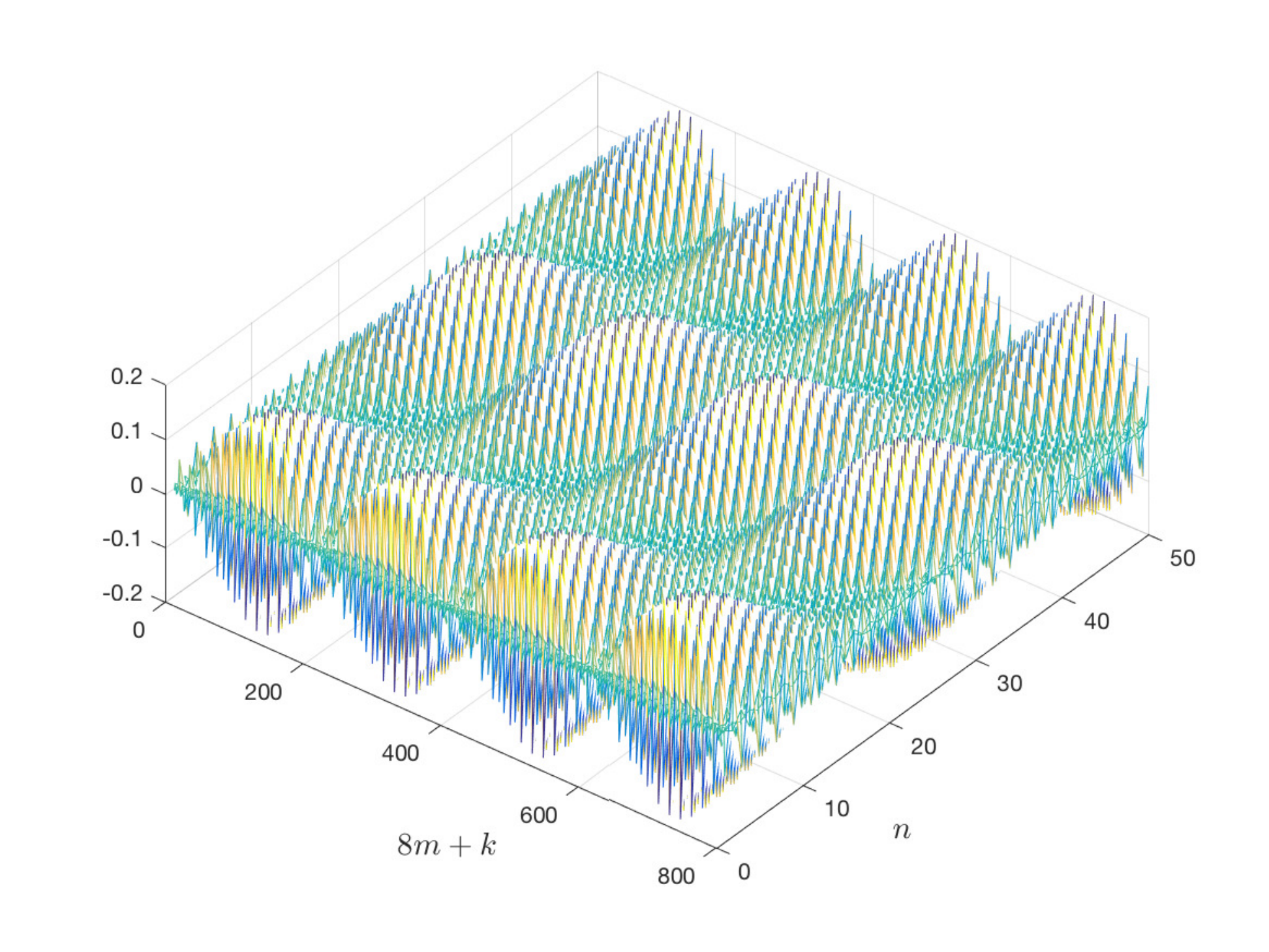}
	\end{subfigure}
	\caption{Evolution of the current flow in 100 cells for $t=\tau+nt_0$ with $n=0,1,...,100$, in the case when $\Gg_1=5$, $\Gg_2=2$, $\Gg_3=1$ and $\Gg_4=11$ and, at $t=\tau$ ($n=0$), we inject a distribution of currents equal to that given by the sum of one of the couples of two conjugate eigenvectors corresponding to the eigenvalues $-0.9980 \pm 0.0628i$ (on the left) and to the eigenvalues $-0.9921 \pm 0.1253i$ (on the right). In particular, the solution on the left is periodic in time and space, but the periodicity is not that of the unit cell.}
	\labfig{check3pha_waves}
\end{figure}

\begin{figure}[!ht]
	\centering
	\includegraphics[width=0.7\textwidth]{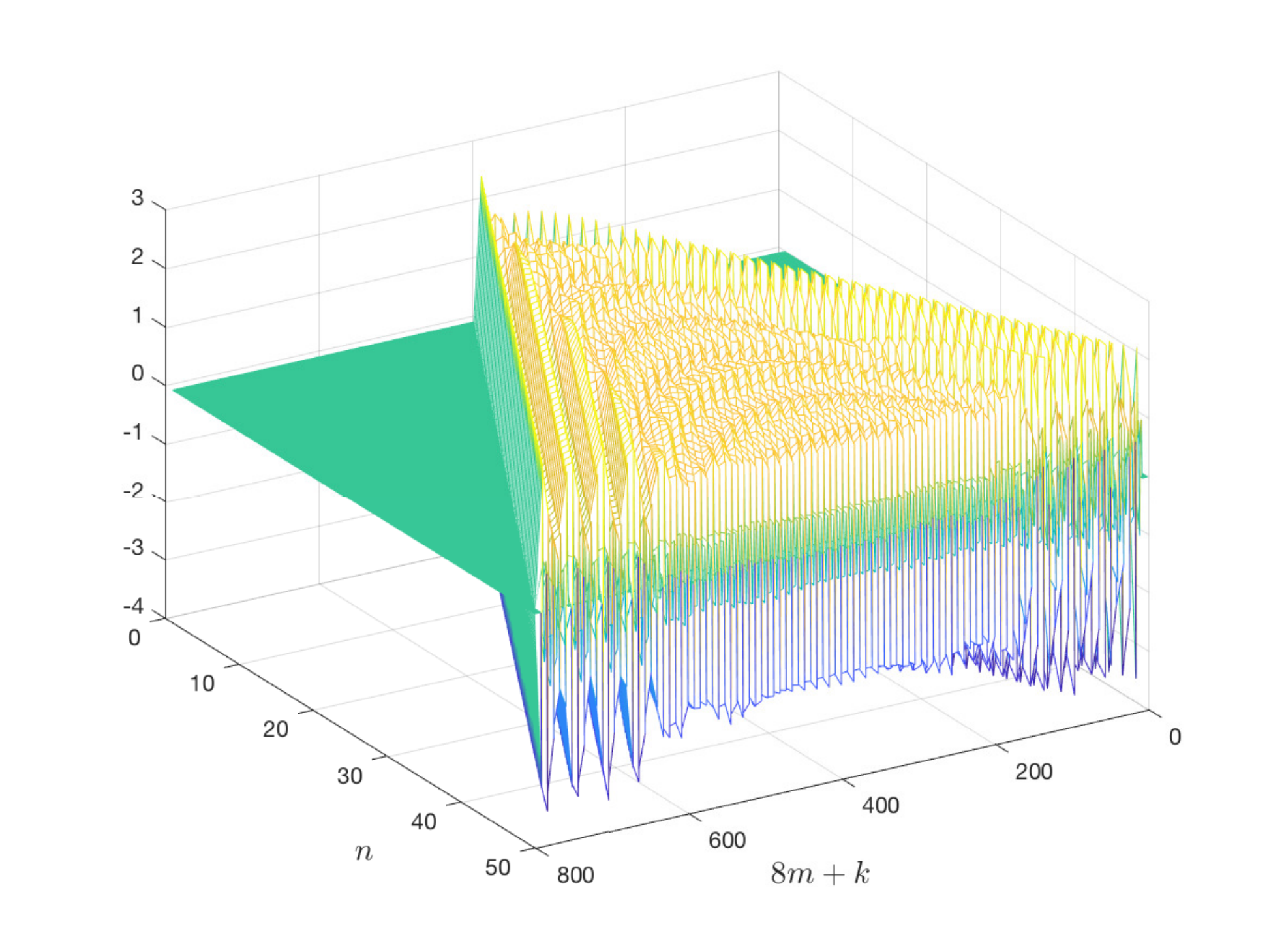}
	\caption{Evolution of the current flow in the space-time checkerboard of \fig{3phase_ratio3}, with $\Gg_1=1$, $\Gg_2=13$, and $\Gg_2=7$, when 100 unit cells are considered (since there are 8 points to monitor in each unit cell, the total number of points is equal to 800). We initially inject a unit current at point 1 of cell 50, that is, at the point labeled with 401, and as time evolves, that is, for $t=\tau+nt_0$ with $n=0,1,\dots,50$, the current flows with the same intensity at the edge of a conic shape having vertex at the point of injection of current. Inside the conic shape the net current is equal to zero,
but is clearly oscillatory and with a non-trivial wake.}
	\labfig{check3phases_inj1}     	
\end{figure}

\section{Concluding remarks}

In this paper we presented some space--time microstructures whose geometry is such that they give rise to field patterns supporting only propagating modes (for a specific range of the material parameters). Accordingly, the transfer matrix that relates the current distribution within the field pattern at a certain time to that after one time period only has eigenvalues on the unit circle. In particular, field patterns supporting only propagating modes occur in two-- and three--phase space--time checkerboards in which the phases all have the same wave speed, independently of the phase impedances, whereas in four--phase checkerboards (with phases having equal wave speed) field patterns support only propagating modes if the impedances of the phases are suitably chosen. Another example of 
field--pattern space--time microstructure is given by a three--phase checkerboard in which the wave speeds of the constituents are in a certain ratio (i.e., $c_2/c_1=c_1/c_3=r$, with $r$ an integer). Specifically, we analyzed two cases: $r=2$ and $r=3$. The first case corresponds to the case of ``broken $\CP\CT$--symmetry'' as, for any choice of the material parameters, field patterns support both propagating modes and modes that blow up (decrease) exponentially with time, whereas for the $r=3$ case, there exists a range of the constituent impedances for which the modes are only propagating modes. 

It is still unclear, either physically or mathematically, why for some space--time microstructures that support field patterns the condition of $\CP\CT$--symmetry is always ``unbroken'' and why for other space--time geometries it could be ``broken'' or ``unbroken'' depending on the choice of the set of material parameters.  Deeper insight is needed.

The very interesting feature related to field patterns without blow-up is that the waves generated by an instantaneous 
disturbance at a point look like shocks with a wake of oscillatory waves. Most remarkably, the amplitude of the oscillations does not tend to zero away from the wave front. It is a new type of wave that we need to better understand.


\section*{Acknowledgements}
The authors are grateful to the National Science Foundation for support through grant DMS-1211359, and for support from
the Institutute for Mathematics and its Applications at the University of Minnesota where this work was begun. They thank Daniel Onofrei for suggesting that it may be worthwhile to look for field patterns in checkerboard geometries.

\section{Appendix A: Green's function for the two-component space-time checkerboard with phases having the same wave speed}\label{Green_function_2phases}

The components of the Green function for the two-component checkerboard illustrated in \fig{check1} take the following expressions when current is injected at

\beq\nonumber
\begin{array}{l}
\text{Point 1 of cell 0}:
\\j(1,0,0)=1
\end{array}\quad\Rightarrow\quad \left\{\begin{aligned} &G(1,1,-1)=1; \,\, G(2,1,-1)=-\frac{\Gg_1-\Gg_2}{\Gg_1+\Gg_2};\,\,G(3,1,-1)=\frac{(\Gg_1-\Gg_2)^2}{2\Gg_1(\Gg_1+\Gg_2)};\\&G(4,1,-1)=\frac{\Gg_1-\Gg_2}{2\Gg_1};\,\,G(2,1,0)=-\frac{\Gg_1-\Gg_2}{\Gg_1+\Gg_2};\,\,G(3,1,0)=\frac{(\Gg_1-\Gg_2)^2}{2\Gg_1(\Gg_1+\Gg_2)};\\&G(4,1,0)=\frac{\Gg_1-\Gg_2}{2\Gg_1}\end{aligned}\right.
\eeq{Green}	
\beq\nonumber
\begin{aligned}
	&\begin{array}{l}
		\text{Point 2 of cell 0}:
		\\j(2,0,0)=1
	\end{array}\quad\Rightarrow\quad \left\{\begin{aligned} & G(3,2,-1)=\frac{\Gg_1-\Gg_2}{2\Gg_1}; \,\, G(4,2,-1)=\frac{(\Gg_1-\Gg_2)^2}{2\Gg_1(\Gg_1+\Gg_2)};\,\,G(1,2,0)=-\frac{\Gg_1-\Gg_2}{\Gg_1+\Gg_2};\\&G(3,2,0)=\frac{\Gg_1-\Gg_2}{2\Gg_1}; \,\, G(4,2,0)=\frac{(\Gg_1-\Gg_2)^2}{2\Gg_1(\Gg_1+\Gg_2)};\,\,G(1,2,1)=-\frac{\Gg_1-\Gg_2}{\Gg_1+\Gg_2};\\&G(2,2,1)=1\end{aligned}\right.
\\&
\\&\begin{array}{l}
	\text{Point 3 of cell 0}:
	\\j(3,0,0)=1
\end{array}\quad\Rightarrow\quad\left\{\begin{aligned} &G(3,3,-1)=1; \,\, G(4,3,-1)=\frac{\Gg_1-\Gg_2}{\Gg_1+\Gg_2};\,\,G(1,3,0)=\frac{(\Gg_1-\Gg_2)^2}{2\Gg_2(\Gg_1+\Gg_2)};\\&G(2,3,0)=-\frac{\Gg_1-\Gg_2}{2\Gg_2};\,\,G(4,3,0)=\frac{\Gg_1-\Gg_2}{\Gg_1+\Gg_2};\,\,G(1,3,1)=\frac{(\Gg_1-\Gg_2)^2}{2\Gg_2(\Gg_1+\Gg_2)};\\&G(2,3,1)=-\frac{\Gg_1-\Gg_2}{2\Gg_2}\end{aligned}\right.
\\&
\\&
\begin{array}{l}
	\text{Point 4 of cell 0}:
	\\j(4,0,0)=1
\end{array}\quad\Rightarrow\quad \left\{\begin{aligned} & G(1,4,0)=-\frac{\Gg_1-\Gg_2}{2\Gg_2}; \,\, G(2,4,0)=\frac{(\Gg_1-\Gg_2)^2}{2\Gg_2(\Gg_1+\Gg_2)};\,\,G(3,4,0)=\frac{\Gg_1-\Gg_2}{\Gg_1+\Gg_2};\\&G(1,4,1)=-\frac{\Gg_1-\Gg_2}{2\Gg_2}; \,\, G(2,4,1)=\frac{(\Gg_1-\Gg_2)^2}{2\Gg_2(\Gg_1+\Gg_2)};\,\,G(3,4,1)=\frac{\Gg_1-\Gg_2}{\Gg_1+\Gg_2};\\&G(4,4,1)=1\end{aligned}\right. 
\end{aligned}
\eeq{Green2}
with $\Gg_1$ referring to the phase colored in white in \fig{check1} and $\Gg_2$ referring to the phase colored in gray. Obviously, all the other components are equal to zero.

\section{Appendix B: Green's function for the three-component space-time checkerboard with phases having the same wave speed}\label{Green_function_3phases}

By suitably replacing the coefficient $\Gg_2$ with $\Gg_3$ (in \fig{3phase_check} phase 1 is white, phase 2 light gray and phase 3 dark gray) in the components of the Green function of the two-phase checkerboard geometry (see Section \ref{Green_function_2phases}), we can recover the components of the Green function for the three-phase checkerboard as follows
\beq\nonumber
\begin{aligned}
	&
\begin{array}{l}
	\text{Point 1 of cell 0}:
	\\j(1,0,0)=1
\end{array}\quad\Rightarrow\quad \left\{\begin{aligned} &G(1,1,-1)=1; \,\, G(2,1,-1)=-\frac{\Gg_1-\Gg_2}{\Gg_1+\Gg_2};\,\,G(3,1,-1)=\frac{(\Gg_1-\Gg_2)^2}{2\Gg_1(\Gg_1+\Gg_2)};\\&G(4,1,-1)=\frac{(\Gg_1+\Gg_2)(\Gg_1-\Gg_3)}{2\Gg_1(\Gg_1+\Gg_3)};\,\,G(2,1,0)=-\frac{\Gg_1-\Gg_3}{\Gg_1+\Gg_3};\\&G(3,1,0)=\frac{(\Gg_1-\Gg_2)(\Gg_1-\Gg_3)}{2\Gg_1(\Gg_1+\Gg_3)};\,\,G(4,1,0)=\frac{\Gg_1-\Gg_2}{2\Gg_1}\end{aligned}\right.
\\&	\\&	
\begin{array}{l}
	\text{Point 2 of cell 0}:
	\\j(2,0,0)=1
\end{array}\quad\Rightarrow\quad \left\{\begin{aligned} & G(3,2,-1)=\frac{\Gg_1-\Gg_2}{2\Gg_1}; \,\, G(4,2,-1)=\frac{(\Gg_1-\Gg_2)(\Gg_1-\Gg_3)}{2\Gg_1(\Gg_1+\Gg_3)};\\&G(1,2,0)=-\frac{\Gg_1-\Gg_3}{\Gg_1+\Gg_3};\,\,G(3,2,0)=\frac{(\Gg_1+\Gg_2)(\Gg_1-\Gg_3)}{2\Gg_1(\Gg_1+\Gg_3)};\\& G(4,2,0)=\frac{(\Gg_1-\Gg_2)^2}{2\Gg_1(\Gg_1+\Gg_2)};\,\,G(1,2,1)=-\frac{\Gg_1-\Gg_2}{\Gg_1+\Gg_2};\,\,G(2,2,1)=1\end{aligned}\right.
\end{aligned}
\eeq{Green_3phases1}
\beq\nonumber
\begin{aligned}
	&
\begin{array}{l}
	\text{Point 3 of cell 0}:
	\\j(3,0,0)=1
\end{array}\quad\Rightarrow\quad\left\{\begin{aligned} &G(3,3,-1)=1; \,\, G(4,3,-1)=\frac{\Gg_1-\Gg_3}{\Gg_1+\Gg_3};\,\,G(1,3,0)=\frac{(\Gg_1-\Gg_2)(\Gg_1-\Gg_3)}{2\Gg_2(\Gg_1+\Gg_3)};\\&G(2,3,0)=-\frac{(\Gg_1+\Gg_2)(\Gg_1-\Gg_3)}{2\Gg_2(\Gg_1+\Gg_3)};\,\,G(4,3,0)=\frac{\Gg_1-\Gg_2}{\Gg_1+\Gg_2};\\&G(1,3,1)=\frac{(\Gg_1-\Gg_2)^2}{2\Gg_2(\Gg_1+\Gg_2)};\,\,G(2,3,1)=-\frac{\Gg_1-\Gg_2}{2\Gg_2}\end{aligned}\right.
\\&	\\&	
\begin{array}{l}
	\text{Point 4 of cell 0}:
	\\j(4,0,0)=1
\end{array}\quad\Rightarrow\quad \left\{\begin{aligned} & G(1,4,0)=-\frac{\Gg_1-\Gg_2}{2\Gg_2}; \,\, G(2,4,0)=\frac{(\Gg_1-\Gg_2)^2}{2\Gg_2(\Gg_1+\Gg_2)};\,\,G(3,4,0)=\frac{\Gg_1-\Gg_2}{\Gg_1+\Gg_2};\\&G(1,4,1)=-\frac{(\Gg_1+\Gg_2)(\Gg_1-\Gg_3)}{2\Gg_2(\Gg_1+\Gg_3)}; \,\, G(2,4,1)=\frac{(\Gg_1-\Gg_2)(\Gg_1-\Gg_3)}{2\Gg_2(\Gg_1+\Gg_3)};\\&G(3,4,1)=\frac{\Gg_1-\Gg_3}{\Gg_1+\Gg_3};\,\,G(4,4,1)=1\end{aligned}\right.
\end{aligned}
\eeq{Green_3phases4}
All the other components are equal to zero. 

\section{Appendix C: Green's function for the four-component space-time checkerboard with phases having the same wave speed}\label{Green_function_4phases}

In this case, the components of the Green function can be obtained from the case of a three-phase checkerboard (see Section \ref{Green_function_3phases}) by suitably substituting the parameter $\Gg_1$ with $\Gg_4$, where phase 1 is the one colored in white and phase 4 the one colored in very light gray in \fig{4phase_check} (phase 2 is gray and phase 3 dark gray). Therefore, we have:
\beq\nonumber
\begin{aligned}
	&\begin{array}{l}
		\text{Point 1 of cell 0}:
		\\j(1,0,0)=1
	\end{array}\quad\Rightarrow\quad \left\{\begin{aligned} &G(1,1,-1)=\frac{(\Gg_1+\Gg_3)(\Gg_2+\Gg_4)}{(\Gg_1+\Gg_2)(\Gg_3+\Gg_4)};\,\, G(2,1,-1)=\frac{(\Gg_1+\Gg_3)(\Gg_2-\Gg_4)}{(\Gg_1+\Gg_2)(\Gg_3+\Gg_4)}; \\&G(3,1,-1)=-\frac{(\Gg_1-\Gg_2)(\Gg_1+\Gg_3)(\Gg_2-\Gg_4)}{2\Gg_1(\Gg_1+\Gg_2)(\Gg_3+\Gg_4)};
		\\&G(4,1,-1)=\frac{-\Gg_2(\Gg_1^2-\Gg_1\Gg_2+3\Gg_1\Gg_3+\Gg_2\Gg_3)+\Gg_4(\Gg_1^2-\Gg_1\Gg_3+3\Gg_1\Gg_2+\Gg_2\Gg_3)}{2\Gg_1(\Gg_1+\Gg_2)(\Gg_3+\Gg_4)};\\&G(1,1,0)=-\frac{(\Gg_2-\Gg_3)(\Gg_1-\Gg_4)}{(\Gg_1+\Gg_2)(\Gg_3+\Gg_4)};\,\,
		G(2,1,0)=-\frac{(\Gg_1-\Gg_3)(\Gg_2+\Gg_4)}{(\Gg_1+\Gg_2)(\Gg_3+\Gg_4)};\\&
		G(3,1,0)=\frac{(\Gg_1-\Gg_2)(\Gg_1-\Gg_3)(\Gg_2+\Gg_4)}{2\Gg_1(\Gg_1+\Gg_2)(\Gg_3+\Gg_4)}
		;\,\,G(4,1,0)=\frac{(\Gg_1-\Gg_2)(\Gg_1+\Gg_3)(\Gg_2+\Gg_4)}{2\Gg_1(\Gg_1+\Gg_2)(\Gg_3+\Gg_4)}\end{aligned}\right.\\&\\
	&\begin{array}{l}
		\text{Point 2 of cell 0}:
		\\j(2,0,0)=1
	\end{array}\quad\Rightarrow\quad \left\{\begin{aligned} &  G(3,2,-1)=\frac{(\Gg_1-\Gg_2)(\Gg_1+\Gg_3)(\Gg_2+\Gg_4)}{2\Gg_1(\Gg_1+\Gg_2)(\Gg_3+\Gg_4)};
		\\&G(1,2,0)=\frac{(\Gg_3-\Gg_1)(\Gg_2+\Gg_4)}{(\Gg_1+\Gg_2)(\Gg_3+\Gg_4)};
		\,\,G(2,2,0)=-\frac{(\Gg_2-\Gg_3)(\Gg_1-\Gg_4)}{(\Gg_1+\Gg_2)(\Gg_3+\Gg_4)};\\&
		G(3,2,0)=\frac{-\Gg_2(\Gg_1^2-\Gg_1\Gg_2+3\Gg_1\Gg_3+\Gg_2\Gg_3)+\Gg_4(\Gg_1^2-\Gg_1\Gg_3+3\Gg_1\Gg_2+\Gg_2\Gg_3)}{2\Gg_1(\Gg_1+\Gg_2)(\Gg_3+\Gg_4)}; \\&G(4,2,0)=-\frac{(\Gg_1-\Gg_2)(\Gg_1+\Gg_3)(\Gg_2-\Gg_4)}{2\Gg_1(\Gg_1+\Gg_2)(\Gg_3+\Gg_4)};
		\,\,G(1,2,1)=\frac{(\Gg_1+\Gg_3)(\Gg_2-\Gg_4)}{(\Gg_1+\Gg_2)(\Gg_3+\Gg_4)};
		\\&G(2,2,1)=\frac{(\Gg_1+\Gg_3)(\Gg_2+\Gg_4)}{(\Gg_1+\Gg_2)(\Gg_3+\Gg_4)};\,\, G(4,2,-1)=\frac{(\Gg_1-\Gg_2)(\Gg_1-\Gg_3)(\Gg_2+\Gg_4)}{2\Gg_1(\Gg_1+\Gg_2)(\Gg_3+\Gg_4)}\end{aligned}\right.\\&\\&
\end{aligned}
\eeq{Green_4phases1}
\beq\nonumber
\begin{aligned}	
&	\begin{array}{l}
		\text{Point 3 of cell 0}:
		\\j(3,0,0)=1
	\end{array}\quad\Rightarrow\quad\left\{\begin{aligned}  &G(3,3,-1)=\frac{(\Gg_1+\Gg_3)(\Gg_2+\Gg_4)}{(\Gg_1+\Gg_2)(\Gg_3+\Gg_4)};\,\, G(4,3,-1)=\frac{(\Gg_1-\Gg_3)(\Gg_2+\Gg_4)}{(\Gg_1+\Gg_2)(\Gg_3+\Gg_4)};
		\\&G(1,3,0)=\frac{(\Gg_1-\Gg_2)(\Gg_1-\Gg_3)(\Gg_2+\Gg_4)}{2\Gg_2(\Gg_1+\Gg_2)(\Gg_3+\Gg_4)};\,\,G(3,3,0)=-\frac{(\Gg_2-\Gg_3)(\Gg_1-\Gg_4)}{(\Gg_1+\Gg_2)(\Gg_3+\Gg_4)};
		\\&G(2,3,0)=\frac{\Gg_2(\Gg_1^2-\Gg_1\Gg_2+3\Gg_1\Gg_3+\Gg_2\Gg_3)-\Gg_4(\Gg_1^2-\Gg_1\Gg_3+3\Gg_1\Gg_2+\Gg_2\Gg_3)}{2\Gg_2(\Gg_1+\Gg_2)(\Gg_3+\Gg_4)};\\&G(4,3,0)=-\frac{(\Gg_1+\Gg_3)(\Gg_2-\Gg_4)}{(\Gg_1+\Gg_2)(\Gg_3+\Gg_4)};\,\,G(1,3,1)=-\frac{(\Gg_1-\Gg_2)(\Gg_1+\Gg_3)(\Gg_2-\Gg_4)}{2\Gg_2(\Gg_1+\Gg_2)(\Gg_3+\Gg_4)};\\&G(2,3,1)=-\frac{(\Gg_1-\Gg_2)(\Gg_1+\Gg_3)(\Gg_2+\Gg_4)}{2\Gg_2(\Gg_1+\Gg_2)(\Gg_3+\Gg_4)}\end{aligned}\right.\\&	\\&
	\begin{array}{l}
		\text{Point 4 of cell 0}:
		\\j(4,0,0)=1
	\end{array}\quad\Rightarrow\quad \left\{\begin{aligned} &  G(1,4,0)=-\frac{(\Gg_1-\Gg_2)(\Gg_1+\Gg_3)(\Gg_2+\Gg_4)}{2\Gg_2(\Gg_1+\Gg_2)(\Gg_3+\Gg_4)};
		\\&G(3,4,0)=\frac{(\Gg_1+\Gg_3)(\Gg_4-\Gg_2)}{(\Gg_1+\Gg_2)(\Gg_3+\Gg_4)};\,\,G(4,4,0)=-\frac{(\Gg_2-\Gg_3)(\Gg_1-\Gg_4)}{(\Gg_1+\Gg_2)(\Gg_3+\Gg_4)};\\&G(1,4,1)=\frac{\Gg_2(\Gg_1^2-\Gg_1\Gg_2+3\Gg_1\Gg_3+\Gg_2\Gg_3)-\Gg_4(\Gg_1^2-\Gg_1\Gg_3+3\Gg_1\Gg_2+\Gg_2\Gg_3)}{2\Gg_2(\Gg_1+\Gg_2)(\Gg_3+\Gg_4)}; \\&G(2,4,1)=\frac{(\Gg_1-\Gg_2)(\Gg_1-\Gg_3)(\Gg_2+\Gg_4)}{2\Gg_2(\Gg_1+\Gg_2)(\Gg_3+\Gg_4)};
		\,\,G(3,4,1)=\frac{(\Gg_1-\Gg_3)(\Gg_2+\Gg_4)}{(\Gg_1+\Gg_2)(\Gg_3+\Gg_4)};
		\\&G(4,4,1)=\frac{(\Gg_1+\Gg_3)(\Gg_2+\Gg_4)}{(\Gg_1+\Gg_2)(\Gg_3+\Gg_4)};\,\, G(2,4,0)=-\frac{(\Gg_1-\Gg_2)(\Gg_1+\Gg_3)(\Gg_2-\Gg_4)}{2\Gg_2(\Gg_1+\Gg_2)(\Gg_3+\Gg_4)}\end{aligned}\right.\\& 
\end{aligned}
\eeq{Green_4phases}
All the other components are equal to zero.

\ifx \bblindex \undefined \def \bblindex #1{} \fi\ifx \bbljournal \undefined
  \def \bbljournal #1{{\em #1}\index{#1@{\em #1}}} \fi\ifx \bblnumber
  \undefined \def \bblnumber #1{{\bf #1}} \fi\ifx \bblvolume \undefined \def
  \bblvolume #1{{\bf #1}} \fi\ifx \noopsort \undefined \def \noopsort #1{} \fi

\end{document}